# Massive young stellar objects in the Local Group irregular galaxy NGC 6822 identified using machine learning

David A. Kinson,[1]⋆ Joana M. Oliveira,[1] and Jacco Th. van Loon[1]
[1] *Lennard-Jones Laboratories, School of Chemical and Physical Sciences, Keele University, ST5 5BG, UK*

**ABSTRACT**

We present a supervised machine learning methodology to classify stellar populations in the Local Group dwarf-irregular galaxy NGC 6822. Near-IR colours ($J-H$, $H-K$, and $J-K$), $K$-band magnitudes and far-IR surface brightness (at 70 and 160 $\mu$m) measured from *Spitzer* and *Herschel* images are the features used to train a Probabilistic Random Forest (PRF) classifier. Point-sources are classified into eight target classes: young stellar objects (YSOs), oxygen- and carbon-rich asymptotic giant branch stars, red giant branch and red super-giant stars, active galactic nuclei, massive main-sequence stars and Galactic foreground stars. The PRF identifies sources with an accuracy of ∼ 90 per cent across all target classes rising to ∼ 96 per cent for YSOs. We confirm the nature of 125 out of 277 literature YSO candidates with sufficient feature information, and identify 199 new YSOs and candidates. Whilst these are mostly located in known star forming regions, we have also identified new star formation sites. These YSOs have mass estimates between ∼ 15 − 50 $M_\odot$, representing the most massive YSO population in NGC 6822. Another 82 out of 277 literature candidates are definitively classified as non-YSOs by the PRF analysis. We characterise the star formation environment by comparing the spatial distribution of YSOs to those of gas and dust using archival images. We also explore the potential of using (unsupervised) t-distributed stochastic neighbour embedding maps for the identification of the same stellar population classified by the PRF.

## 1 INTRODUCTION

Resolved star formation has been extensively studied on large scales in both the Milky Way and Magellanic Clouds (MCs). Stepping out to a distance of ∼ 490 kpc (e.g. Sibbons et al. 2012, 2015) NGC 6822 is the closest dwarf irregular galaxy to the Milky Way beyond the MCs. With no known companions (see for example De Blok & Walter 2000), and no previous interactions with large Local Group galaxies M31 or the Milky Way (McConnachie et al. 2021), NGC 6822 presents itself as a non-tidally disrupted analogue to the SMC. By understanding how star formation progresses in NGC 6822 the impact of tidal interactions on triggering star formation can be better constrained. Understanding massive star formation in a metal poor environment has implications for studies of the early universe as well, and NGC 6822 provides an analogue for typical star forming galaxies at $z \sim 2$.

NGC 6822 has a metallicity approximately equal to that of the SMC (∼ 0.2 $Z_\odot$, e.g. Skillman et al. 1989; Richer & McCall 2007) and it is relatively gas-rich with very conspicuous large scale East-West 'wings' of H I gas (Volders & Högbom 1961). The total H I mass is estimated to be $1.38 \times 10^8$ $M_\odot$ (Mateo 1998), and the molecular and dust masses are respectively $M_{\rm mol} < 1 \times 10^7$ $M_\odot$ (Gratier et al. 2010) and $M_{\rm dust} = 2.9^{+2.8}_{-0.8} \times 10^5$ $M_\odot$ (Rémy-Ruyer et al. 2015). Using these mass estimates, Schruba et al. (2017) find a gas-to-dust ratio of $480^{+170}_{-240}$. The total stellar mass is $1.5 \times 10^8$ $M_\odot$ (Madden et al. 2014), giving a observed baryonic mass of ∼ $2.9 \times 10^8 M_\odot$. Weldrake et al. (2003) find a total dark matter mass to 5 kpc (the extent of the H I disk) of ∼ $3.2 \times 10^9$ $M_\odot$, implying that NGC 6822 is heavily dark-matter-dominated.

The H I gas distribution in NGC 6822 has a very intricate structure. It is dominated by a large under-density or cavity seen to the South-East of the main galaxy body (e.g. Gottesman & Weliachew 1977; De Blok & Walter 2000). The inner rim of this cavity is edged by optical emission that could be linked to its origin in large-scale stellar feedback (Cannon et al. 2012), although no agreement has been reached on the mechanism responsible (De Blok & Walter 2000). Opposing this feature on the North-West wing of the main H I distribution there is a large over-density of gas. It has been suggested that this over-density is due to the presence of a putative interacting companion (e.g. De Blok & Walter 2000, 2003). However this hypothesis is not supported by stellar population studies across NGC 6822 (Cannon et al. 2012). A likely explanation for the complex extended H I structure in NGC 6822 is a warped disk inclined with respect to the line of sight (e.g. Cannon et al. 2012).

Clearly apparent in NGC 6822 is the central bar which runs nearly perpendicular to the H I gas distribution in a North-South direction for ∼ 1.4 kpc (∼ 10 arcmin; see Fig. 1). This central bar is host to the young stellar component of the galaxy, with older populations more elliptically distributed (e.g. Letarte et al. 2002; Hirschauer et al. 2020). The central bar is boxed at either end by bright star forming regions (SFRs) first identified by Hubble (1925) with ages up to 10 Myr (Efremova et al. 2011; Bianchi et al. 2012). Attempts to find sites of star formation beyond the bar, namely in the H I over-dense region, have so far been unsuccessful (Schruba et al. 2017) despite promising indicators in the distribution of H I gas (De Blok & Walter 2000, 2003).

⋆ Email: d.a.kinson@keele.ac.uk

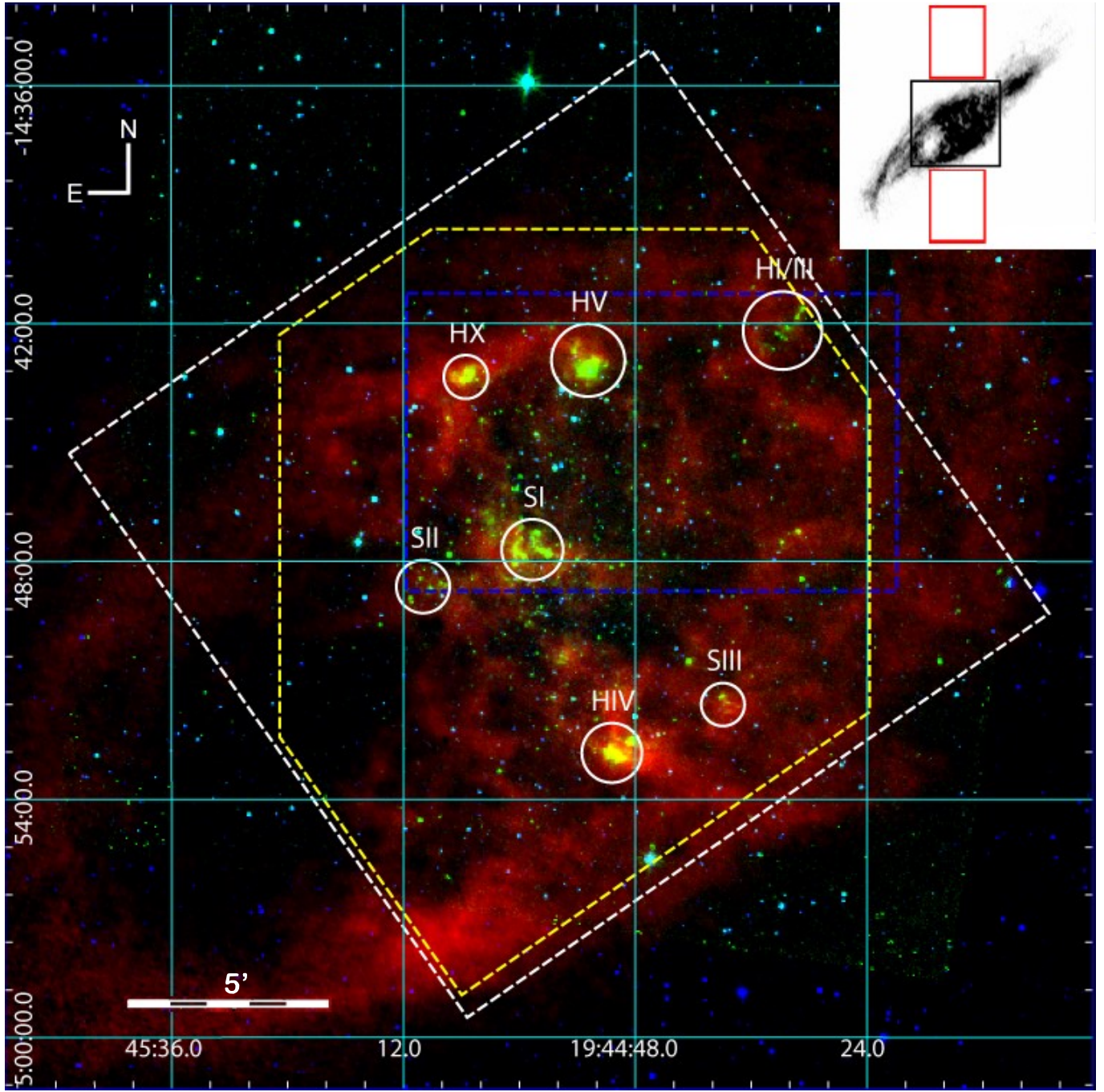

**Figure 1.** An RGB image of NGC 6822 showing H I gas emission (red, Schruba et al. 2017), 8 $\mu$m *Spitzer* IRAC (green, Kennicutt et al. 2003) and 2MASS *K*-band[1] (blue) images. The area covered by this study is shown by the dashed yellow line. The coverage of the far-IR *Herschel* PACS images is given by the white dashed line. CO (2−1) coverage from Gratier et al. (2010) is shown by the blue dashed rectangle. Major SFRs are identified. The cavity in the H I distribution can be seen in the lower left of the image. Note the H I coverage extends far beyond the area of the main image, see inset upper right. The off-galaxy fields used for Galactic foreground comparison in Sect. 4.2.2 and Appendix B are indicated by the red outlines in the inset H I image.

CO emission is often used as a proxy for molecular hydrogen (which does not emit at radio wavelengths) due to their general spatial coincidence. No CO maps of the entirety of the central bar of NGC 6822 have yet been produced, with published studies focusing on the brightest SFRs (Gratier et al. 2010; Schruba et al. 2017). Schruba et al. (2017) produced ALMA high-resolution maps of several small (110 × 110 arcsec$^2$) fields in CO (2–1), four of which are centred on the most prominent SFRs: Hubble I/III, IV, V and X. They find CO cores with typical sizes of ∼ 2.3 pc and propose that such small scales could be the cause of the low levels of CO emission seen in many dwarf galaxies, due to poor beam filling at lower resolutions.

---

[1] https://old.ipac.caltech.edu/2mass/

Previous studies of resolved young stellar object (YSO) populations in NGC 6822 on a galaxy wide basis have used established colour-cuts (Jones et al. 2019) or basic statistical (Hirschauer et al. 2020) classification criteria. In Jones et al. (2019) candidate YSOs were found using a series of mid-infrared (mid-IR) colour-magnitude diagram (CMD) cuts developed by Whitney et al. (2008) and Sewiło et al. (2013). The spectral energy distributions (SEDs) of those candidates were fitted initially using stellar atmosphere models (Castelli & Kurucz 2003) in order to remove contaminant objects. The sources remaining were then compared to YSO model grids (Robitaille et al. 2006; Robitaille 2017). Sources were assigned to one of three confidence levels based on the goodness of the fit to the best-fit model and colour-cuts criteria.

In addition to the four well-known SFRs already mentioned (Hubble I/III, IV, V and X), Jones et al. (2019) studied in detail three other significant SFRs, which they label Spitzer I, II and III (their table 9 provides the positions and sizes). To the South of Spitzer I lie regions identified in Hubble (1925): Hubble VI and VII, a young open star cluster (Chandar et al. 2000) and a globular cluster (Huxor et al. 2013) respectively, while Spitzer II borders Hubble IX, a cluster of undetermined age (Huxor et al. 2013). Jones et al. (2019) remove from their YSO candidate lists any sources within the half-light radius of the globular cluster Hubble VII. Spitzer I is particularly prominent with an infrared excess noted by Cannon et al. (2006) and CO (2–1) emission identified by Gratier et al. (2010). This region seems to be more active in terms of star formation than the other optically brighter Hubble regions (Jones et al. 2019).

Using the same near-infrared (near-IR) and mid-IR catalogues, Hirschauer et al. (2020) applied colour cuts developed using kernel density estimate techniques to separate different stellar populations. YSO candidates were identified based on consistent CMD positions as well as being located within one of the SFRs discussed in Jones et al. (2019). The major SFRs were all recovered in the resulting YSO distribution, however fewer YSOs were identified compared to Jones et al. (2019) due to different limiting magnitude cuts applied to the classifications.

A more holistic method to classify YSOs, which does not rely on potentially imperfect models or a piece-wise approach and takes into account interdependancies and degeneracies between observable features, is therefore needed; this provides the motivation for using machine learning techniques. The paper is organised as follows. Section 2 presents the archival data used in our analysis. Section 3 introduces the machine learning methods used, the results of which are given in Section 4. Section 5 discusses the resulting YSO catalogue and examines the star forming environments in NGC 6822. Finally in Section 6 we conclude by summarising our analysis.

## 2 DATA

### 2.1 Near-IR Data

#### 2.1.1 Near-IR catalogues

We used the near-IR aperture photometry catalogue from Sibbons et al. (2012), constructed from Wide Field Camera (WFCAM) images obtained on the United Kingdom Infrared Telescope (UKIRT Casali et al. 2007). The focal plane array of WFCAM is comprised of four Rockwell Hawaii-II detectors (Casali et al. 2007). To fill in the gaps between the detectors four exposures are required, resulting in a tile image covering 0.8 deg². Several tiled images were used to construct the catalogue of Sibbons et al. (2012). A single tile image is large enough to cover the major star forming regions in NGC 6822. We retrieved the *JHK* images from the WFCAM Science Archive (WSA), fully processed using the standard WFCAM pipeline by CASU[2], that will be used to perform new aperture photometry as described in the next section. Full details on the data acquisition, reduction and catalogue generation can be found in Sibbons et al. (2012).

This catalogue contains ∼ 375,000 sources over an area of 3 deg$^2$ centred on NGC 6822. The catalogue is estimated to be complete to depths of $J$ = 19.5 mag and $K$ = 18.7 mag (Sibbons et al. 2012). Our analysis is restricted to the extent of the star forming bar in part due to the smaller field of view in the far-IR images (illustrated in Fig. 1), but more importantly as this is where star formation activity is occurring (e.g. Letarte et al. 2002; Hirschauer et al. 2020), thus allowing us to validate our new methodology for YSO identification. These requirements give a total area considered in this paper of approximately 0.07 deg$^2$ containing ∼ 15,000 near-IR sources.

In addition, our analysis also makes use of 328 sources in or behind the Magellanic Clouds. These sources are included for the purposes of training our supervised classifier, where comparable data is not available for NGC 6822. Near-IR ($JHK_s$) photometric data for these sources was obtained as part of the MCs survey conducted using the SIRIUS camera on the InfraRed Survey Facility (IRSF) at the South African Astronomical Observatory; full details of the data aquisition and reduction, as well as catalogue construction are reported in Kato et al. (2007). IRSF photometry was transformed onto the WFCAM photometric system using the conversions detailed in Appendix C.

#### 2.1.2 New near-IR aperture photometry

Upon close inspection of both catalogue and images it was apparent that the catalogue did not include aperture photometry of point-sources towards the central regions of the bright SFRs. Given that the goal of our analysis is YSO identification we extracted our own aperture photometry. As already mentioned, we used the fully processed images retrieved from the WSA.

Aperture photometry was performed using the PHOTUTILS package for PYTHON (Bradley et al. 2020). We used 3.57 pixel apertures, the WSA standard radius. Apertures were placed at the known *Spitzer* position for each Jones et al. (2019) source, and the new aperture photometry was calibrated using ∼ 9000 sources with near-IR photometry in the Sibbons et al. (2012) catalogue. We found a 1-$\sigma$ dispersion of 0.055 mag or less in each band between our photometry and that in the published catalogue for these calibration sources.

This process recovered near-IR magnitudes for an additional 54

---

[2] https://research.ast.cam.ac.uk/vdfs/documentation.html#wsystem

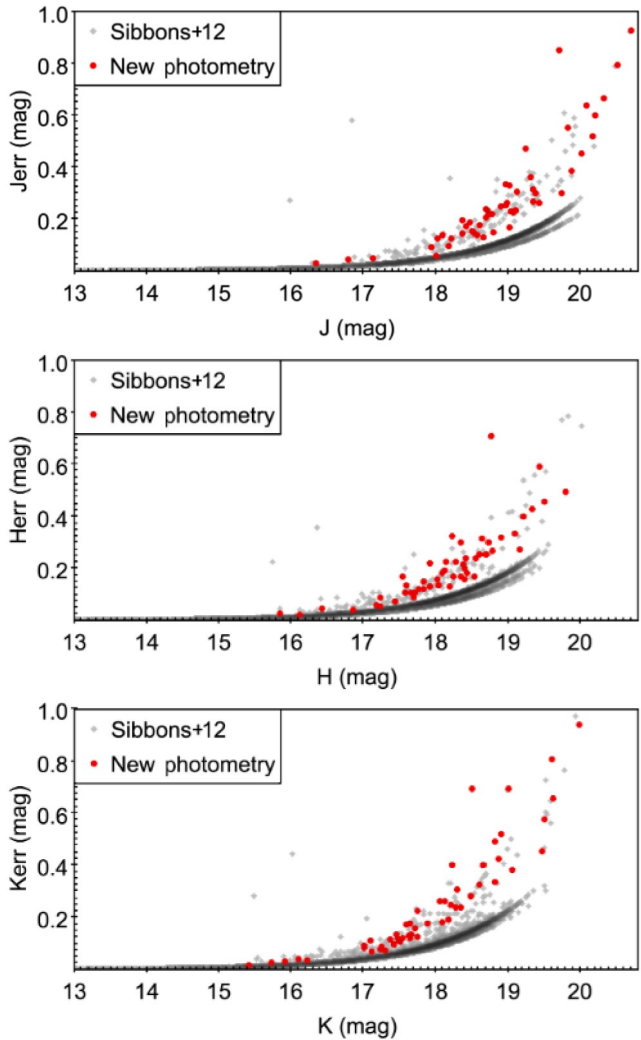

**Figure 2.** Magnitudes and uncertainties of the new near-IR aperture photometry compared to the catalogue of Sibbons et al. (2012). We refer the reader to the source paper for any data issues in that catalogue. Red circles indicate those *Spitzer*-identified sources in NGC 6822 for which we have performed aperture photometry.

sources located in the coverage gaps (bright SFRs) in the Sibbons et al. (2012) catalogue which were added to the photometric catalogue. Properties of the final catalogue are shown in Fig. 2. The new photometric uncertainties are at the higher end of the range of values seen in the extant near-IR data, reflecting the high background levels encountered in these bright SFR.

## 2.2 Large aperture far-IR photometry

In addition to the near-IR data we also include far-IR data in our analysis. The goal is to provide a feature for the classifiers which indicates proximity to a wider star forming environment. Far-IR emission can be used to trace UV light from young stellar populations which is reprocessed by surrounding dust (Bianchi et al. 2012).

Galametz et al. (2010) obtained far-IR images of NGC 6822 using the *ESA Herschel Space Observatory* (Pilbratt et al. 2010). The PACS (Poglitsch et al. 2010) images at 70 and 160 $\mu$m were retrieved from the ESA *Herschel* Science Archive [3].

Magellanic point sources of known characteristics are used as part of the training set for the machine learning (see Sect. 3.4.6) therefore similar far-IR measurements to those performed in NGC 6822 were needed. We used 70 $\mu$m *Spitzer* MIPS (Rieke et al. 2004) and 160 $\mu$m *Herschel* PACS (Poglitsch et al. 2010) images of the SMC and LMC (Meixner et al. 2006; Gordon et al. 2011; Meixner et al. 2013). Inspection of the Magellanic 160 $\mu$m images in particular revealed the presence of a small (non-astrophysical) bias level; corrections were applied to bring the brightness zero level of each image into line with one another, full details are provided in Appendix D.

At the position of each *K*-band source in the near-IR catalogues we performed a simplified aperture photometry on the far-IR images using a large radius. We take the sum of the image counts within this aperture in both 70 and 160 $\mu$m using the same physical radius for each and average for the number of image pixels within this aperture. An aperture radius equivalent to 30 pc around each source was chosen based on the typical scales from theoretical predictions of infrared dark cloud sizes (Tan et al. 2014), and comparison with the CO emission tracing dust in NGC 6822 (Schruba et al. 2017). We adopt distance moduli of $\mu = 23.34$ mag for NGC 6822 (Jones et al. 2019) and of 18.49 and 18.90 mag for the LMC and the SMC respectively (Pietrzyński et al. 2013; Hilditch et al. 2005); this results in apertures sizes equivalent to ∼ 12.7 arcsec at the distance of NGC 6822, and ∼ 103 and 124 arcsec respectively at the distances of the SMC and the LMC. These measurements are included in Table 1 for the PRF training set.

## 3 MACHINE LEARNING METHODS

We use both supervised and unsupervised methods to classify the sources in NGC 6822 using their observed properties. In this section we outline the basic principles of the specific methods used as well as providing details of the data on which the methods are run.

### 3.1 Unsupervised methods: t-distributed stochastic neighbour embedding (t-SNE)

Unsupervised methods in machine learning refer to those which operate with minimal human intervention in the processing of unlabelled data. This allows previously unknown relations in data to be found and can also be useful in classification of data where labels may be unreliable. Any relations found by such unsupervised methods arise entirely from the data and therefore are not potentially biased by input classifications as in supervised machine learning.

There are many unsupervised machine learning approaches to data classification. We use the t-distributed Stochastic Neighbour Embedding (t-SNE) method, a type of self organising map (Van der Maaten & Hinton 2008). A t-SNE produces a 2D map of higher dimensional data in which sources of similar properties are often clustered.

This is achieved by firstly turning the provided catalogue data into joint probabilities between data entries. The technique then attempts to minimise the divergence between the two-dimensional positions of the entries and their higher dimensional counterparts. This results in a 2D map where each individual source is clustered nearby to those with similar higher dimensional characteristics. As a result of this method of higher dimensionality plotting the t-SNE is not able to handle missing data.

---
[3] http://archives.esac.esa.int/hsa/whsa/

There are several fine tuning parameters which can affect the outcome of the map such as the number of iterations in the calculations and the number of neighbouring sources compared to in each calculation of the divergence difference, known as the perplexity value.

We use the SKLEARN t-SNE implementation for PYTHON (Pedregosa et al. 2011) to create the maps as this allows for simple control of the iteration and perplexity parameters. Key limitations of the t-SNE method are the prohibitively large run time for large catalogues and the memory limitations of the system to hold the large volume of high dimensional information during the calculation process. Therefore, for any given dataset there is a 'sweet spot' where good separation of the data is reached without excessive run time. In our analysis we set the perplexity and number of iterations values to 200 and 500 respectively. The t-SNE maps are discussed in Sect. 4.5.

## 3.2 Supervised methods

Supervised machine learning is a useful tool in situations where a relationship in one set of data can be applied to a second set with the same measurements but unknown classification. In machine learning these measurements are called features and the classifications targets or target classes. Often in astronomy supervised machine learning involves training on a set of sources with known properties and applying this to another set of data in which a specific object class is of interest, e.g. evolved stars in Hernandez et al. (2021) or YSOs in Cornu & Montillaud (2020). One of the best established supervised classifiers is the Random Forest Classifier (RFC, Breiman 2001).

### 3.2.1 Random forest classifiers (RFC)

An RFC in its simplest form is a set of randomised decision trees which each return a classification for each source in the data. At each node in the tree a threshold value for a set of features is implemented which splits the decision path for the data input. This is repeated over a large number of randomly generated trees. The majority decision amongst all the trees is then given as the RFC classification for each object.

The RFC is trained on a subset of 'known' sources against which its classification accuracy can be measured. This is achieved by splitting the training data set into a sample for training and a sample to test against. Most commonly this is done randomly/pseudo-randomly with a random seed; the latter of which is the method used in this work (Sect. 4.1). The accuracy of the classifier on the test set can be assessed at this stage to provide a measure of the classifier's performance.

The accuracy of the classifier is inherently linked to the quality of the training data in both the extent of the feature parameter space covered by the objects in each class in the training set and the similarity of the training set to the data to be classified.

A final consideration is the reliability of the classifications in the training set. The more sources with incorrect target classes in the training set the worse the RFC will perform in classification. This can be minimised via conservative construction of the training set samples (Sect. 3.4) and using a random forest classifier which can account for uncertainties in the training data target class labelling. A probabilistic random forest classifier offers this feature, as well as being able to account for uncertainties in feature data.

### 3.2.2 Probabilistic random forest (PRF)

A Probabilistic Random Forest (PRF) is a variation on the traditional RFC approach which takes uncertainties into account in both the features and training class labels. The uncertainty in each feature is used to build a probability distribution which is taken into account at each node of the decision trees. The probability of each path is split at each node based on the probability distribution in that feature, rather than using a threshold condition against which each source is judged; an illustration of this is shown in fig. 1 of Reis et al. (2019). Furthermore by modelling class labels as probability functions and propagating these through its trees a PRF has been shown to outperform a classical RFC in classification accuracy where class labels may be impure (Reis et al. 2019).

Unlike a classical RFC model which requires all feature data to be present, if feature information is missing in the data for a given node the PRF can propagate to the next nodes on an even split basis. In this way a source with incomplete data can be classified which is not possible in a classical RFC, even if such a classification is obviously less reliable. For these obvious advantages we use a PRF rather than a classical RFC in this work. An in-depth description of the PRF implementation can be found in Reis et al. (2019).

## 3.3 Classification features

Our machine learning methods (supervised and unsupervised) were trained on six features: near-IR $K$-band magnitude, three near-IR colours ($J-H$, $H-K$ and $J-K$) and two far-IR brightnesses at 70 and 160 $\mu$m (see Sects. 2.1 and 2.2).

We classify sources using the PRF on a minimum of two out of four near-IR features. This allows for one missing band from the near-IR $JHK$ data, for example a missing $H$-band value would affect two near-IR features $J-H$ and $H-K$. Those sources which lack the sufficient features were removed from the catalogue. Sources which presented clear issues in many features such as un-physical colours (e.g. $J-K \ll 0$) or excessive error bars (e.g. $(J-K)_{err} > 1$ mag) were also removed. In total $\sim 2.5$ per cent of the sources in the original near-IR catalogue were removed. This left a catalogue of 11,341 sources remaining.

## 3.4 Sources in the training set

The training set for the PRF was constructed from various extant catalogues containing sources in eight target classes: YSOs, Oxygen-rich Asymptotic Giant Branch stars (OAGBs), Carbon Asymptotic Giant Branch stars (CAGBs), Red Giant Branch stars (RGBs), Red Super-Giant stars (RSGs), Active-Galactic Nuclei (AGNs), Massive Main Sequence stars (MMSs) and Galactic Foreground stars (FGs). The observed properties of sources in each of these classes are shown in the CMD, colour-colour diagram (CCD) and far-IR brightness plots in Fig. 3.

As previously noted, the performance of the classifier is linked to the numerical size of the data-set, how much parameter space each class samples and the labelling accuracy of the training data. To ensure the highest reliability of test sample target labels we included only sources identified in the literature using methods other than broad-band photometry, e.g. spectroscopy, narrow-band indices or *Gaia* proper motions. The training set sources in NGC 6822 are matched to the near-IR catalogue of Sibbons et al. (2012) using a 1 arcsec search radius. A summary of the information for each training set class is provided in Table 2. For some classes the number of sources is relatively small, however the parameter space occupied by that class is often small and thus the sampling remains good (see Fig. 3). It is important to note that given that there is not an 'unknown' class, all sources in the catalogue must be assigned to one of the training set classes. This will inevitably lead to classification contamination, which we discuss in Sect. 4.2.2. The individual classes in the training set are described in detail below.

**Table 1.** Positions, measurements and their uncertainties (where available) as well as source classification from the training set. A single row for each training set class is shown here, the full version is available in the supplementary material. Near-IR magnitudes are presented in the WFCAM photometric system.

| RA (J2000) h:m:s | Dec (J2000) deg:m:s | $J$ mag | $J_{err}$ mag | $H$ mag | $H_{err}$ mag | $K$ mag | $K_{err}$ mag | [70] MJy sr$^{-1}$ | [160] MJy sr$^{-1}$ | Target Class |
|---|---|---|---|---|---|---|---|---|---|---|
| 05:04:51.69 | −66:38:07.4 | 18.83 | 0.043 | 18.02 | 0.040 | 17.07 | 0.034 | 21578.2 | 138989.8 | YSO |
| 19:44:34.63 | −14:55:52.0 | 17.52 | 0.036 | 16.61 | 0.024 | 16.41 | 0.026 | 1818.6 | 22382.3 | OAGB |
| 19:44:32.41 | −14:56:30.8 | 17.86 | 0.047 | 16.85 | 0.029 | 16.27 | 0.023 | 12720.6 | 600.6 | CAGB |
| 00:37:04.67 | −73:22:29.6 | 18.05 | 0.050 | 17.29 | 0.070 | 16.47 | 0.060 | 405.8 | 837.7 | AGN |
| 19:44:26.62 | −14:56:38.2 | 16.90 | 0.022 | 16.55 | 0.022 | 16.43 | 0.028 | 0 | 0 | FG |
| 19:44:47.45 | −14:54:28.9 | 19.09 | 0.131 | 18.21 | 0.094 | 18.10 | 0.107 | 5316.9 | 35.2 | RGB |
| 19:44:55.70 | −14:51:55.9 | 13.26 | 0.003 | 12.58 | 0.002 | 12.36 | 0.002 | 9917.9 | 44576.6 | RSG |
| 19:45:03.02 | −14:54:27.1 | 18.00 | 0.053 | 17.95 | 0.075 | 18.21 | 0.117 | 401.7 | 7192.1 | MMS |

**Table 2.** Information on the eight target classes included in our training set. The classification method and reference are given, as well as the number of sources in each class. The AGN sample are identified using a variety of methods (see Sect. 3.4.5). More details of all these classes are provided in Sect. 3.4.

| Class | Number of Sources | Identification Method |
|---|---|---|
| YSO | 43 | *Spitzer*-IRS spectra |
| OAGB | 99 | VIS/NIR spectra, narrowband indices |
| CAGB | 461 | VIS/NIR spectra, narrowband indices |
| AGN | 89 | Various |
| FG | 500 | *Gaia* proper motions |
| RGB | 124 | *Spitzer*-IRS spectra |
| RSG | 192 | Optical & *Spitzer*-IRS spectra |
| MMS | 18 | *Gaia* proper motions |

### 3.4.1 Asymptotic giant branch stars

We include a well defined AGB training set as these stars can have similar near-IR colours and magnitudes to massive YSOs (see Fig. 3). The AGB training samples consist solely of previously classified sources in NGC 6822. Most of the AGB sources originate from Sibbons et al. (2012) identified initially with near-IR photometry and further confirmed with spectroscopy (Sibbons et al. 2015). Additional AGB sources come from the four-band catalogue (*R*, *I*, CN and TiO) from Letarte et al. (2002) and the spectroscopic catalogue from Kacharov et al. (2012), which utilises low-resolution VIMOS spectroscopy to confirm AGB nature. We distinguish between O- and C-rich AGBs which present different colours due to their distinct atmospheric molecular composition. We use these classifications to create two AGB target classes in our training set. The training data includes 560 AGBs, split between 461 CAGBs and 99 OAGBs; this difference in class size does not significantly affect the PRF's training since they occupy distinct and reasonably compact regions of parameter space, as shown in Fig. 3.

### 3.4.2 Red giants and supergiants

Red giant and supergiant stars are two different populations which contaminate YSO samples at opposite ends in terms of magnitude. Red supergiants are a bright, dusty (similar to AGBs) and young population which may be located close to sites of recent star formation. Red giant branch (RGB) stars are an older, more dynamically evolved population that tends to be more smoothly distributed over the body of a galaxy (see for example Cioni et al. 2000, for the SMC) and therefore are less likely to be tightly correlated with sites of far-IR emission. Whilst RGB stars are rarely dusty (Van Loon 2008) they will likely contribute significantly to the YSO contaminants towards our sensitivity limit which is ∼ 1.4 mag below the tip of the RGB (K = 18.11 mag, Hirschauer et al. 2020).

The training sample for RGBs come from three spectroscopic catalogues. A sample of RGBs in Local Group dwarf galaxies are used in Kirby et al. (2013) to constrain the galaxies' metallicities, by determining the Fe/H ratio from spectroscopy; we include the NGC 6822 RGBs in our training set. The catalogues of Tolstoy et al. (2001) and Swan et al. (2016) both use spectra containing the Ca II triplet centred at 850 nm to quantify the metallicity ratios in RGB stars. This training class contains 124 sources.

The RSG class for the training set is drawn from catalogues in NGC 6822 and the LMC. In NGC 6822 the spectroscopically confirmed samples of Massey (1998) and Massey et al. (2007) include 22 sources. Given this small number, we augment it by including LMC RSG sources, from the catalogues of Jones et al. (2017) which are based on *Spitzer*-IRS spectroscopy, as well as some additional sources from Neugent et al. (2020) identified with spectroscopy focused on Balmer and TiO lines from 340 nm to 1 $\mu$m. The training class contains a total of 170 LMC RSGs, giving a total of 192 sources.

### 3.4.3 Foreground Galactic sources

To define a training class of foreground Galactic contaminants in our analysis we started by crossmatching the *Gaia* EDR3 catalogue[4] (Gaia Collaboration et al. 2020) with the near-IR data (1 arcsec matching radius). This recovered 5007 near-IR sources with *Gaia* counterparts. Subsequently proper motion (PM) measurements were employed to identify high-reliability Galactic contaminants (see also Sect. 3.4.4 for similar analysis for MMS stars). We analysed the PM components in right ascension (RA) and declination (Dec) separately.

---
[4] https://www.cosmos.esa.int/web/gaia/earlydr3

We select sources with PM(RA) outside the range −3 to 3 mas yr$^{-1}$; for PM(Dec) the equivalent limits are −5 to 3 mas yr$^{-1}$. These values were selected based on the distribution of PM measurements on and off the main body of NGC 6822 (see Appendix B). These cutoff values were intentionally conservative; a sample of 500 foreground sources is identified and included in our training set. Any remaining foreground sources without reliable *Gaia* PMs will be classified by the machine learning processes. We discuss the foreground training and recovered sets further in Sect. 4.2.2.

### 3.4.4 Massive main sequence stars

Massive main-sequence stars in NGC 6822 come from the catalogue of Bianchi et al. (2001). We took the bluest sources ($B - V < 0.4$ mag) as suggested by these authors. We further apply a near-IR cut ($J - K < 0.2$ mag), based on the intrinsic colours of O- and B-type stars (Zombeck 2006) and the average reddening estimates of $E(B - V) = 0.35$ mag (Bianchi et al. 2001).

This sample was then matched to the *Gaia* EDR3 catalogue to obtain PMs in a similar process as for the FG class above. To select sources in NGC 6822 only, we set PM limits both for the RA and Dec components between −2 and 2 mas yr$^{-1}$ (more details in Appendix B). Using these *Gaia* PM measurements provides an additional level of certainty beyond the broad band photometric selection.

The near-IR catalogue samples the brightest main-sequence sources, therefore we identify only nineteen sources for this class. This is the smallest class in the training set; Sect. 4.2.1 shows the effect this has on the training process.

### 3.4.5 Active galactic nuclei

Active Galactic Nuclei (AGN) are also known contaminants of YSO samples (e.g. Whitney et al. 2008; Sewiło et al. 2013; Jones et al. 2017) and their near-IR colours show considerable overlap (see Fig. 3). The large aperture far-IR measurements are thus crucial to differentiate between YSOs which are strongly correlated with nearby far-IR emission and AGN which as background objects have no such preferential correlation on large scales. The very recent update of the MILLIQUAS compilation (the Million Quasars Catalog, version 7.2, Flesch 2021) does not include any spectroscopically confirmed AGN in our field of analysis. Therefore we choose for the training set AGNs located behind the SMC; this sample will be analysed in detail in Pennock et al. (in preparation). It was compiled from a variety of surveys employing different methods including: Magellanic Quasars Survey (Kozłowski et al. 2011); MACHO Spectroscopy (Geha et al. 2003), *Chandra* observations and OGLE optical to near-IR photometry (Dobrzycki et al. 2003), *XMM-Newton* and *WISE* mid-IR photometry (Maitra et al. 2018) as well as VLT/FORS2 spectra (Ivanov et al. 2016). Near-IR photometry for these sources originates from the IRSF catalogue of Kato et al. (2007) and was converted to the WFCAM system using the transformations given in Appendix C. We have a total of 89 sources with sufficient feature data which are taken into the training set.

### 3.4.6 Young stellar objects

As previously discussed YSO candidates have been identified within SFRs in the central bar of NGC 6822 (Jones et al. 2019; Hirschauer et al. 2020). These analyses were based on *Spitzer* colour cuts and/or SED fitting. We required additional confirmation of an object's nature therefore we do not immediately include these samples in the training set. Instead the initial YSO training set was constructed from spectroscopically confirmed YSOs in the SMC (Oliveira et al. 2013) and LMC (Jones et al. 2017): massive YSOs from embedded Stage I sources to more evolved ultracompact H II regions that are unresolved using *Spitzer* in the MCs. The spectroscopic classification of these YSOs relies mostly on *Spitzer*-IRS spectra, and uses a variety of spectral features in the $5-20\,\mu$m range.

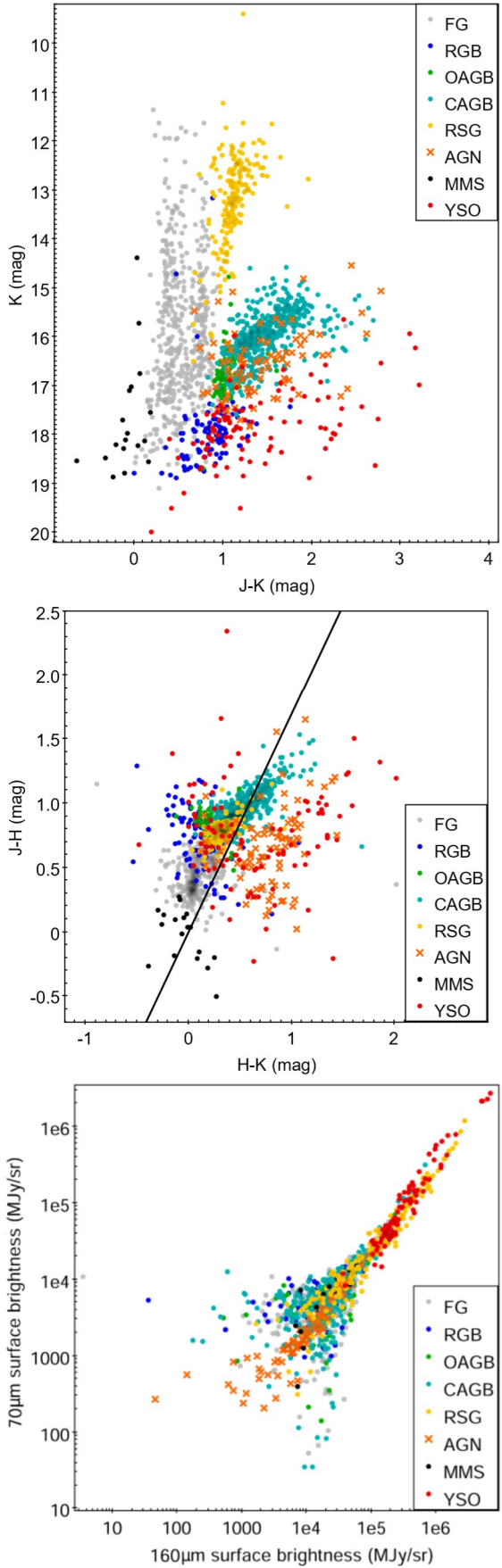

**Figure 3.** CMD, CCD and far-IR brightness plots of the sources in our initial training set. The reddening line shown in the CCD is calculated from the values given in Rieke & Lebofsky (1985).

After conversion to the WFCAM photometric system (see Appendix C), the magnitudes of the Magellanic YSOs were scaled to the distance of NGC 6822. Furthermore these sources were selected such that they are brighter than the detection threshold for the NGC 6822 data of $K = 19.5$ mag (Sibbons et al. 2012). In total 43 MC YSOs are included in the training set, 39 from the LMC and four from the SMC. These MC YSOs are by design amongst the most massive, but are well-matched to the sample that can be identified with the present near-IR survey. In Sect. 4.1.2 we further compare these sources to YSO candidates in NGC 6822.

### 3.4.7 Exclusion of planetary nebulae from classification

YSO samples can also be contaminated by planetary nebulae (PN). We considered the PNe candidates from the analysis of Leisy et al. (2005) which surveyed a large area in NGC 6822, a seventeen-strong sample that the authors state is complete down to 3.5 mag below the brightest PN. However, only one PN candidate has a near-IR counterpart. Therefore our near-IR catalogue seems in fact too shallow to detect all but the very brightest PN in NGC 6822.

Nevertheless, we further considered the samples of Leisy et al. (1997) and Jones et al. (2017) in the LMC. This resulted in 29 PN which would be detectable in our near-IR catalogue when shifted to the distance of NGC 6822. Upon closer inspection we found that these LMC PNe are of rare types (e.g. proto-PN), or their PNe nature is questionable or ambiguous. Introducing these sources into the training set would lead to a significant bias in the classifier towards potentially rarer or uncertain types. Furthermore, taking into account the stellar mass of the LMC and NGC 6822, respectively $2.7\times10^9\,M_\odot$ (Besla 2015) and $1.5\times10^8\,M_\odot$ (Madden et al. 2014) few such objects would be expected in NGC 6822. This reinforces our conclusion above that very few if any PNe are present in our near-IR catalogue, and therefore a PN class is not needed in the training set. For completeness we note that the single PN with a near-IR counterpart is classified as an AGN by the PRF classifier (Sect.4).

## 4 RESULTS

### 4.1 Initial PRF outcomes

With the training set defined for our eight target classes we ran the PRF on the remaining catalogue data. The classifier was run 20 times with different random seeds for the test/train splitting to eliminate any stochastic effects in training data selection. This splitting is done on a global rather than class-wise basis, therefore leading to some unevenness in testing data class sizes (see Sect. 4.1.1). This was done on a 75 per cent training, 25 per cent test split, which provides a robust sample to train on in all target classes even those with a low total number of sources in the training set data.

This method is somewhat similar to a k-fold cross-validation approach to training classifiers (for a theoretical introduction see Mosteller & Tukey 1968). In our method however we include all features in each PRF run rather than excluding one per fold as a way of estimating feature importance and classifier performance. This was done as we have a relatively small number of features, in principle all equally important. This was established from testing with t-SNE maps with individual features removed, see Sect. 4.5. Applying the PRF in multiple runs risks encountering issues associated with overfitting of the data especially in small target classes. Given the accuracies returned (Sect. 4.1.1) for each class we do not believe overfitting to be an issue in this application.

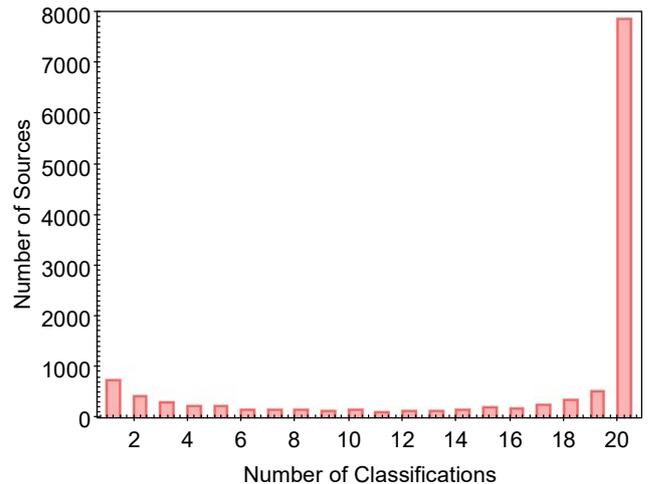

**Figure 4.** A histogram of the PRF classifications across the eight classes and twenty runs. Most sources (∼ 79 per cent) are consistently classified in the same class ($n_{\rm class}=20$).

For each PRF run we generate a list of source classifications as well as a set of confusion matrices (Sect. 4.1.1). Using the ACCURACY_SCORE function in SKLEARN each run returns an estimated accuracy of correct classification across all classes. For the 20 runs of the PRF this varies from 84 to 91 per cent.

For every source we obtain a value $n_{\rm class}$; this is the number of runs in which a source is classified into the given class and this value allows us to assess the confidence for the object to belong to each particular class. For the training data, most test sources are consistently classified into the same (correct) class. Due to the random nature of the train/test sampling, each source in the training set is effectively classified a different number of times, and therefore it is not meaningful to assign them global $n_{\rm class}$ values. For the classification of the rest of the catalogue ∼ 79 per cent of sources are identified consistently into the same target class over all 20 runs (Fig. 4). This is indicative of a robust classification system which is independent of biases induced by random sampling effects and the sources included in the training set. It also shows that the classifier is able to effectively distinguish between target classes.

### 4.1.1 Confusion matrices

A confusion matrix is a standard tool in supervised machine learning which provides a measure of the accuracy of a classifier in its application to the training data prior to wider application on unclassified data. Each matrix shows the statistics of the actual and predicted labels for the test sources. A confusion matrix for a perfect classifier will show a diagonal of 100 per cent accurate classifications. In practice however some classes may perform better than others; the matrices identify those classifications that are the most significant cause of confusion.

For each run of the classifier two matrices are generated: one with the raw number of sources for each class and one which is normalised

MMSs that are classed as FG, a likely consequence of the similarities in near-IR colours. Such misidentifications are not seen in reverse (i.e. FG to MMS) suggesting that this effect is exacerbated by the small number of MMS sources in the test portion of the training set (Fig. 3).

FG sources are well recovered, with a small level of confusion into the RGB class. A greater number of RGBs are incorrectly identified as FG sources. This occurs at fainter magnitudes beyond the depth at which *Gaia* counterparts could be found (see Sect. 4.2.2 for further discussion).

The YSO class does not suffer from any contamination from other classes (see first column of the matrices in Fig. 5). YSO misclassifications occur into the CAGB and AGN classes (top row of the matrices in Fig. 5). Fig. 3 shows that these classes overlap significantly in near-IR features with YSOs. The inclusion of the far-IR features in the classification scheme clearly added discriminating power, reducing any confusion to low levels, ∼ 11 per cent for both classes. The matrix values shown in Fig. 5 are representative of all seeds, with significant variations occurring only for runs in which the sampling of a particular class is poor.

### 4.1.2 Extending the YSO training set

The promising results from the initial PRF runs, with very successful classification for MC YSOs (Figs. 4 and 5), motivated the application of the PRF classifier to the near-IR counterparts of YSO candidates in the catalogues of Jones et al. (2019) and Hirschauer et al. (2020), with the intention of confirming their nature and expanding the training set.

YSOs identified in Jones et al. (2019) are split into three confidence tiers, which properties we summarise below. All sources have high CMD scores, suggesting that their colours are consistent with a YSO nature. The 105 high-confidence YSOs further have low reduced-$\chi^2$ fits to YSO models (Robitaille et al. 2006; Robitaille 2017). The 88 medium-confidence YSOs have SEDs relatively poorly fit by YSO models. Finally there are 555 lower-confidence YSOs classified in Jones et al. (2019). These sources were excluded from their SED fitting analysis due to insufficient mid-IR data points, a disjointed SED or indication of a stellar photosphere; some of these sources may still be bona-fide YSO candidates but their nature could note be appropriately constrained. Of these three types, respectively 23, 18 and 195 have counterparts in our near-IR catalogue.

Hirschauer et al. (2020) focuses on identifying a variety of dusty stellar populations in NGC 6822 and therefore does not provide YSO confidence levels in the same way as Jones et al. (2019). Hirschauer et al. (2020) identify 310 YSO candidates, 59 of which are unique from those classified in Jones et al. (2019). Of these 59 unique sources 41 have a near-IR counterpart.

Whilst the PRF is capable of classifying a source with missing features, as described in Sect. 3.3, the quality of these classifications will be reduced owing to the increased number of nodes in each tree at which an even split rather than a probabilistic decision is made. From the YSO candidates of all confidence levels in the Jones et al. (2019) and Hirschauer et al. (2020) catalogues there were 277 sources out of 807 for which we had enough features for the PRF to make a meaningful classification.

The PRF classifies many of these 277 sources with a high level of certainty: as shown in Fig. 6 the $n_\text{class} \geqslant 19$ bins contain 40 per cent of the sources. Of the 277 sources, 82 were classified as YSOs for some of the PRF runs, 55 of which have $n_\text{YSO} \geqslant 19$. Of these 55 sources, 47 are from the tables of Jones et al. (2019) with 10, 4 and 33 coming from their high, medium and lower reliability classifications by the variable number of sources in each test class (Fig. 5). The un-normalised matrices allow us to track any potential imbalances between the number of sources in each training set classes, while the normalised matrices provide an easy to interpret measure of classification accuracy for each target class.

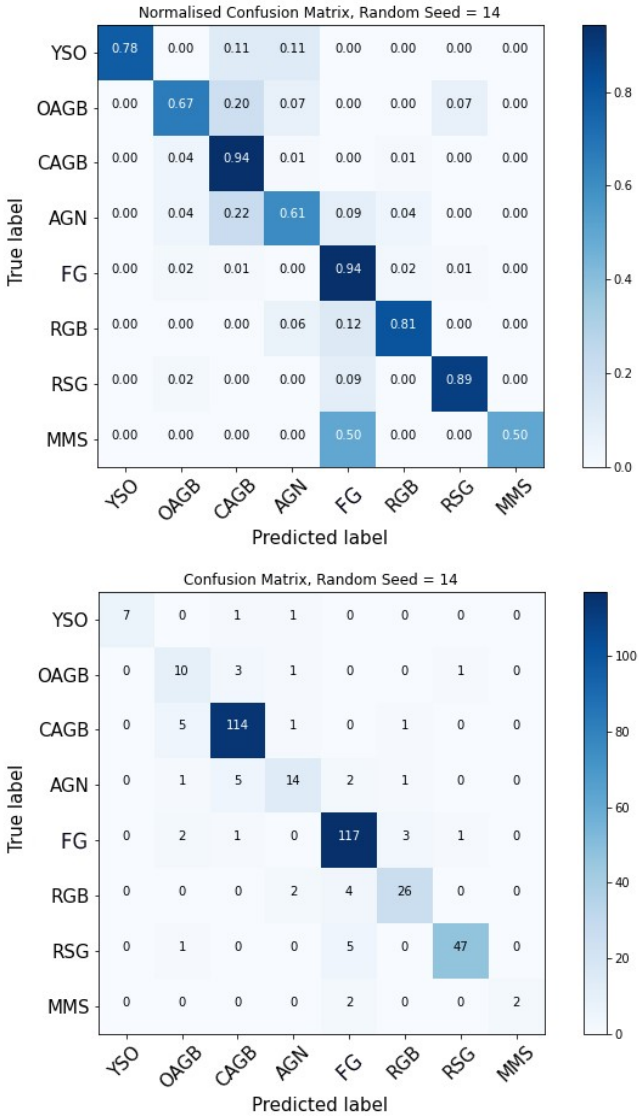

**Figure 5.** A normalised (top) and un-normalised (bottom) confusion matrix for a single run of our PRF classifier using the initial training set. Both matrices were generated from the run with random seed = 14.

In the normalised confusion matrices we see a high rate of correct identification for most classes. Issues arise only between sources of similar observed properties such as OAGBs and CAGBs: for instance the dustiest OAGBs are sometimes classed as CAGBs due to the similarities in their colours. Additionally for both AGB classes there is some confusion with RSGs and AGNs due to the fact that fainter RSGs can have similar features to massive AGB stars (Fig. 3). Some AGNs have SEDs that peak at mid-IR wavelengths and thus can exhibit IR colours similar to AGB stars (Hony et al. 2011; Van Loon & Sansom 2015), and spatially are also uncorrelated with large-scale far-IR emission; therefore some classifier confusion between these classes is not unexpected.

We see the highest degree of misclassification of any class for the respectively. This represents 43 per cent of the highest confidence YSOs from Jones et al. (2019) used in our analysis. The remaining eight are sources unique to the YSO classifications of Hirschauer et al. (2020). These 55 sources are added to the training set, boosting the number of YSOs from 43 to 98. The PRF classifier was retrained on this extended training set for its application across our full NGC 6822 catalogue. The remaining 222 literature YSO candidates which do not meet the threshold of $n_{\rm YSO} \geqslant 19$ are included in the catalogue for the PRF classification; their final classifications are discussed in Sect. 5.1.1.

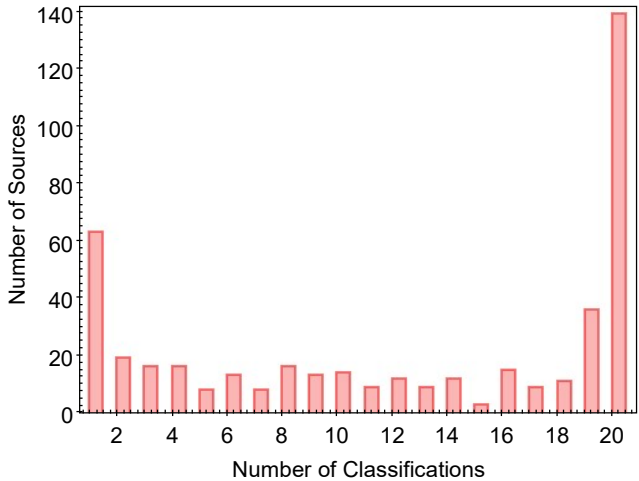

**Figure 6.** A histogram of the $n_{\rm class}$ values across all eight target classes for the literature YSOs from Jones et al. (2019) and Hirschauer et al. (2020) considered for extension of the training set.

The YSOs originating from the MCs are on average redder than those from NGC 6822 (Fig. 7). This is unsurprising given that the MC sample was selected for $5-30\,\mu$m *Spitzer*-IRS spectroscopy, from which their classification is derived. The bottom panel of Fig. 7 shows the far-IR brightnesses for the YSOs in both NGC 6822 and the MCs. The plot includes the loci of dusty sources at various temperatures (adopting a dust emissivity coefficient $\beta = 1.5$), generated using 1D blackbody models in A­stropy (Price-Whelan et al. 2018). All YSOs generally follow the locus for a dust temperature of $T \sim 25$ K. There may be a slight hint that the MC YSO far-IR brightnesses could be consistent with a marginally lower dust temperature. This would be expected given the differences in metallicity between the LMC (from which most MC YSOs originate) and NGC 6822, as metallicity and dust temperature have been shown to be anti-correlated (Van Loon et al. 2010). However, such effect if present seems modest.

This enhanced YSO training set covers a wider region in parameter space for all the used features; furthermore it provides a training set that is now dominated by sources in NGC 6822, mitigating any potential issues relating do differing YSO properties in these galaxies.

### 4.2 Enhanced PRF classifier

The PRF classifier with the extended YSO training set was trained and applied 20 times for the classification of the full catalogue. This was done with the same 20 seeds to determine the split in train/test data as used for the original PRF runs, allowing us to assess the improvement in classification directly. The range of accuracy scores for these new runs is between 87 and 92 per cent. This is a minor improvement overall, however by comparing the normalised confusion matrices it is clear that for some classes (including YSOs) the improvement is more pronounced.

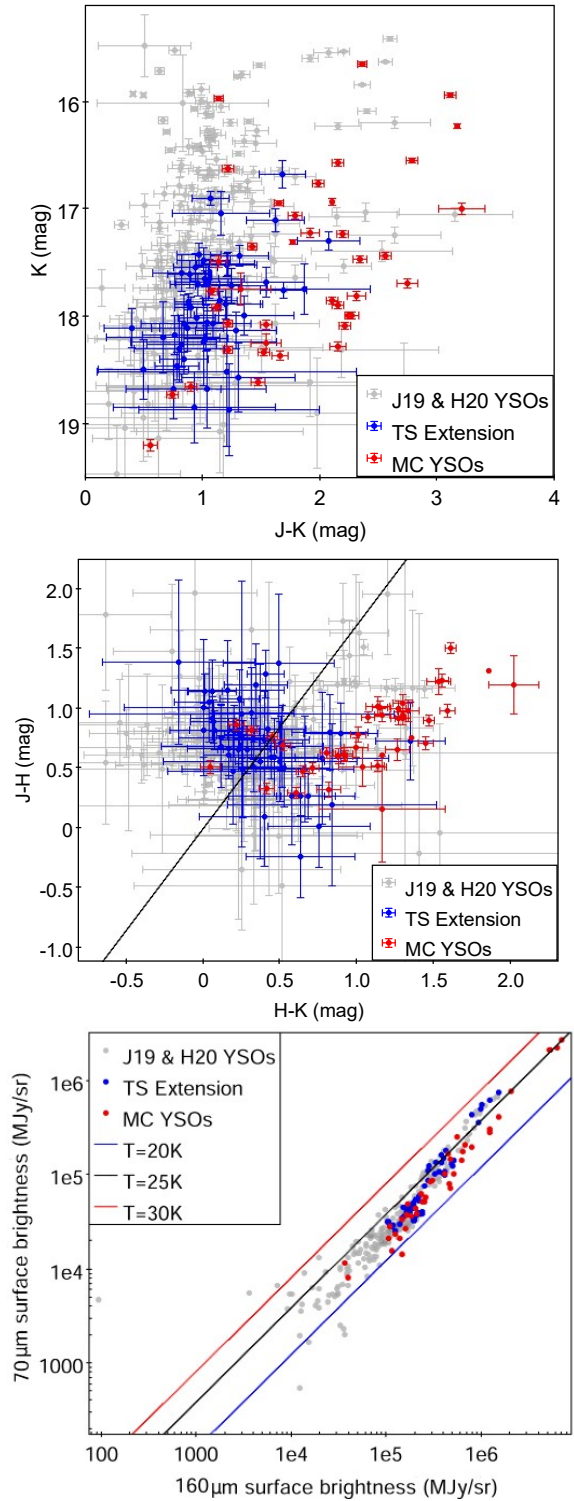

**Figure 7.** CMD (top), CCD (middle) and far-IR brightness plot (bottom) of the YSOs considered for the training set (TS) extension. The YSOs from Jones et al. (2019, J19) and Hirschauer et al. (2020, H20) are shown in grey. The sources identified for inclusion in the extension of the YSO training set are shown in blue. Training set YSOs from the MCs (red circles) have been scaled to the distance of NGC 6822. The reddening line in the CCD is the same as that in Fig. 3. In the far-IR brightness plot (bottom) theoretical loci for dusty blackbodies at various temperatures are shown.

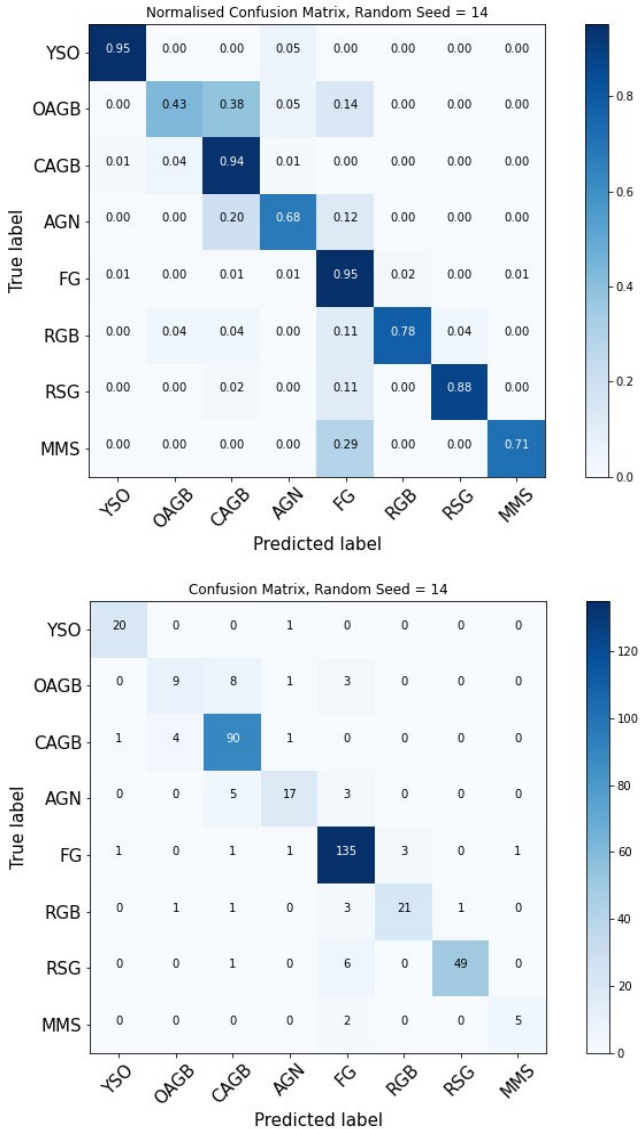

**Figure 8.** A normalised (top) and un-normalised (bottom) confusion matrix for a run of our PRF classifier using the extended training set. The random seed used is the same as that for Fig. 5.

*4.2.1 New confusion matrices*

In the same manner as the initial runs, confusion matrices were generated for each PRF run. The example normalised confusion matrix in Fig. 8 shows that the PRF identifies well all classes except AGN and MMS in the training data. A clear boost in the rate of correctly classified YSOs can be seen by comparing the values in Figs. 5 and 8: an increase from 78 per cent to 97 per cent. Some misclassification of AGN sources remains. FG sources are confused for sources in NGC 6822 only in a very small number of cases. The asymmetric confusion between FG and RGB classes is still present (further discussion in Sect. 4.2.2).

We present all the confusion matrices both normalised and un-normalised generated in our PRF runs using the extended training set in Figs. E1 and E2 for completeness. Across all classes and runs the PRF has a predicted average accuracy of 90 per cent, with class to class variations, exceeding 96 per cent for YSOs.

*4.2.2 Galactic foreground estimation*

As previously discussed both for the initial and improved runs of the PRF, a confusion between FG and RGB sources is seen in the training/test data. We investigate this confusion in our final classification and compare this to foreground models.

TRILEGAL (Girardi et al. 2005) was used to estimate the predicted number of foreground sources for the same area on the sky as covered by the near-IR catalogue, using the detection limit at $K = 19.5$ mag. The modelled foreground suggests that there should be $\sim 2978$ Galactic sources above this threshold.

The PRF runs classify a total of 3082 sources as Galactic foreground in one or more runs, with 2511 classified with $n_{FG} = 20$. Taking only the most certain Galactic sources, those from the training set and those classified with $n_{FG} = 20$, we obtain 3011 foreground contaminants compared to the 2978 predicted by TRILEGAL.

Due to the limiting depth of the *Gaia* data (that corresponds to $K \sim 18.6$ mag), the FG training set is restricted to brighter magnitudes. This limitation is reflected in the FG classifications by the PRF which drop off rapidly below $K \sim 17.5$ mag (Fig. 9). As seen from the confusion matrices, misclassified FG sources are often classed as RGB. We compared the number of output FG and RGB sources to the TRILEGAL model and a Northern off-field region of equal area to our target field (more details in Appendix B) in Fig. 9. We focus on a range of colours centred on the vertical CMD sequence in the foreground data in which the confusion with RGBs is expected to be more prevalent, $0.6 \leqslant J-K \leqslant 0.9$ mag. There are 1537 $n_{RGB} = 20$ sources and 1237 $n_{FG} = 20$ sources within this colour range.

The number of $n_{FG} = 20$ sources closely matches what is seen in the off-field data (with a slight excess compared to the TRILEGAL model predictions) down to $K \sim 17.5$ mag. At fainter magnitudes, the number of model sources continues to grow, overtaking the detected sources as the completeness limit is reached. Below 17.5 mag, as the number of FG sources drops off sharply, the RGB class indeed begins to dominate.

The comparison above confirms that in this colour range a significant number of faint FG sources are misclassified as RGB sources. Using Fig. 9, we estimate that the contamination in this colour and magnitude range of the RGB class by Galactic sources is $\sim 54$ per cent. Taking 54 per cent of the $n_{RGB} = 20$ sources within the colour range $0.6 \leqslant J-K \leqslant 0.9$ mag in addition to the most certain FG sources gives a total estimate of 3840 Galactic foreground sources.

In the off-target field in the colour range where FG sources are classified by the PRF ($0.2 \leqslant J-K \leqslant 0.9$ mag, see Fig. 10), there are 3877 sources. This agrees remarkably closely with the estimated Galactic foreground from our classification once the RGB class contamination is accounted for. Our estimated foreground, while higher, is consistent with the TRILEGAL simulation which uses an approximate parameterised model for the Galaxy (Girardi et al. 2005).

**4.3 Comparing classifier outputs to the training set**

In this section we compare the properties of the PRF classified sources and the training data. Fig. 10 shows the CMD, CCD and far-IR brightness plots for the sources which are always classified in the same class in all runs ($n_{class} = 20$); this figure can be directly compared to Fig. 3 for the training set.

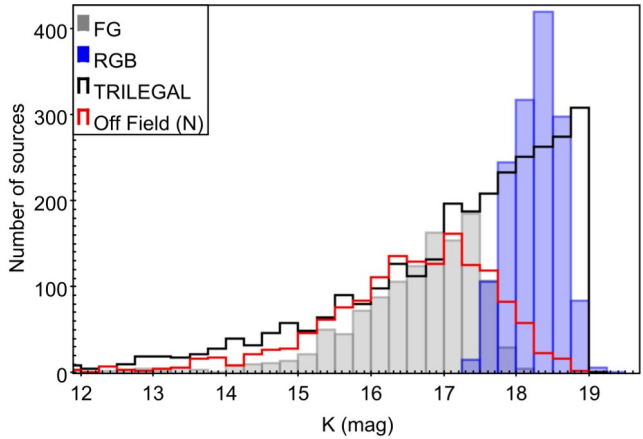

**Figure 9.** A histogram of *K*-band magnitudes for $n_{\rm class} = 20$ FG and RGB sources in the colour interval $0.6 \leqslant J{-}K \leqslant 0.9$ mag. Foreground estimates from the Northern off-target field and TRILEGAL are indicated.

Even though all classes occupy similar positions in the individual diagrams for the training and output data sets, there are however some differences. The RGB class, whilst fairly sparsely populated in the training set plots (Fig. 3), is the most numerous class in the PRF's output (Fig. 10). As we discussed in the previous section, all faint FG sources are misclassified in this class.

Furthermore, as discussed in Sect. 3.4 each source must be classified into one of the eight target classes. In the training set, RGB sources occupy a region of near-IR parameter space shared by the bulk of sources in the NGC 6822 catalogue with no strong relation to far-IR emission. Hence a source with no extreme features is likely to be classified by the PRF model as an RGB star.

The third panel of Fig. 10 shows which classes have the strongest association with the far-IR emission. YSOs are the dominant class with very high far-IR values. RSG are also relatively young and as such are still expected to be spatially associated with far-IR emission; indeed the RSGs' brightnesses extend to higher than average values. Confusion between these two classes is however unlikely, since RSGs are bright and well separated in the CCD from the YSOs (see also the matrices in Fig. 8).

### 4.4 Spatial distributions

In Fig. 11 we show the spatial distribution of sources for each target class: sources with $n_{\rm class} = 20$ and those for which the given class is the largest but $n_{\rm class} < 20$ are colour coded. We highlight a few salient qualitative points below, however a detailed study of the spatial distribution of classes other than YSOs and young RSGs is beyond the scope of this paper.

The AGB sources show a decrease in number with increasing galactocentric radius and, while globally correlated with the known galactic structure, appear less constrained to the central bar compared to classes of younger stellar populations (RSG and YSO). This is consistent with the roughly spheroidal distribution of AGB stars described by Letarte et al. (2002) and seen in Hirschauer et al. (2020).

RGB sources are fairly evenly distributed across the field, the source density gradient between central and outer regions apparently consistent with a population older and more dynamically evolved than the AGB classes (as seen also in e.g. SMC, Cioni et al. 2000). We identify very few RGB sources inside the major SFRs, likely due to increased crowding that makes such regions less complete to faint

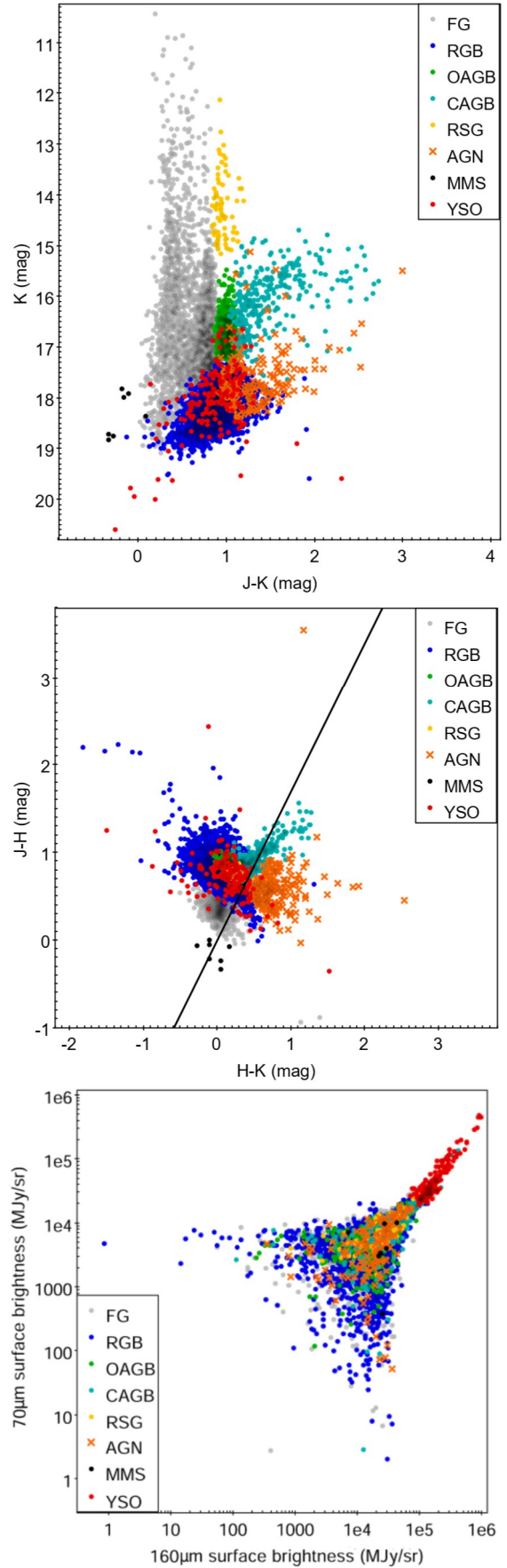

**Figure 10.** CMD, CCD and far-IR brightness plot of the $n_{\rm class} = 20$ sources from the improved PRF classification.

RGB stars. It is important to note that we do not see any contamination between YSO and RGB classes in our confusion matrices (Fig. 8).

The distribution of MMS sources is difficult to comment on due to the low number of sources in this classAs expected FG and AGN sources are evenly distributed across the field. The FG distribution shows a weak hint of the galaxy bar; as shown also from the confusion matrices (see Sect. 4.2.1 and Fig. 8) the FG is expected to be contaminated by RSGs and RGBs at the brighter end. We discuss the spatial distribution of RSGs in Sect. 4.4.1 and the distribution of the YSOs in Sect. 5.1.2.

### 4.4.1 RSG distribution

The RSGs classified by the PRF represent a young ($\gtrsim 10$ Myrs) population in NGC 6822. The locations of the most certain RSGs ($n_{RSG} = 20$) to some extent trace past star formation in the galaxy.

The RSGs occupy the bar of the galaxy filling in between the major SFRs indicated by the YSOs (Fig. 12). This is in agreement with existing models of the star formation history in NGC 6822 which suggest a bar-centric burst of star formation in the last 200 Myr (De Blok & Walter 2000). The current SFRs could be evolutionarily linked to this slightly older population, we discuss the relative ages of the SFRs in Sect. 5.2.

An additional spur in the RSG distribution is seen to the South-East of the bar; this feature is present but not discussed in fig. 12 of Hirschauer et al. (2020). This region borders the large cavity in the H I emission (e.g. De Blok & Walter 2000) and has been linked to both far-UV (Bianchi et al. 2012) and H$\alpha$ (Massey et al. 2007) emission suggesting young populations are present. The far-IR emission is modest, suggesting lack of dust reprocessing, and furthermore we do not classify any YSOs in this region. The detection of RSGs could help tighten estimates of the age of this H I feature, if as discussed in De Blok & Walter (2000) its origin is linked to a burst of previous star formation.

## 4.5 t-SNE maps

Using all features in our PRF catalogue, a t-SNE map of the data was created to compare the class separations in a purely data driven, unsupervised machine learning application. Including both far-IR brightnesses, which trace different temperature dust, improves the separation of target classes in parameter space. A worsening in the class separations was observed when any of the near-IR colours or the $K$-band magnitude were removed.

Figure 13 shows the t-SNE maps from the the training set and catalogue classified by the PRF. The training set sources and PRF classification outputs with $n_{class} = 20$ are shown respectively on the left- and right-hand panels, colour-coded according to the target class. In both maps it can be seen that some classes are tightly grouped (e.g. CAGB) and others are spread over large areas (e.g. RGB). Whilst the area covered by the RGB class is similar in both maps the sparseness of the training set over its parameter space is clear when compared to the similarly sized training class for OAGB sources which are distributed over a smaller map area. The AGB classes occupy a spur to the bottom of the diagram in both the training and output classifications. The CAGB class in particular is very well separated from other classes, suggesting that the use of a t-SNE map with these features could be a useful tool for future studies of evolved stars.

The RSG sources in the training set are tightly packed in a spur off the lobe occupied by the FG sources (see below) and are well separated from AGB sources in the training set. However, there are some newly classified OAGB sources in the same area in the output map. This reflects the inherent difficulties in separating dim RSGs and bright OAGBs in regions of the parameter space where these classes naturally overlap (Fig. 10): even the spectra of such sources are often similar (see sect. 5.2 of Jones et al. 2017, and references therein).

YSOs are located in several tightly defined clumps. In the training set the vast majority of YSOs occupy an island at the bottom-right of the map, the remaining YSOs are scattered outside of the lobe dominated by FG sources, with some sources seen at the tip of the AGB spur. In the $n_{class} = 20$ map, the main island of YSOs from the training set map is partially recovered, and smaller groupings of YSOs are also co-located with scattered YSOs in the training set map. The YSOs seen at the tip of the AGB spur in the training set map are not present in the $n_{class} = 20$ map. We know that CAGBs are some of the reddest sources in our catalogue, and the YSO training set contains objects that are redder than the vast majority of YSOs recovered by the PRF (see Figs. 3 and 10). The lack of YSO sources in this area in the $n_{class} = 20$ map could be a reflection of this difference.

FG sources in the training set occupy the lobe to the lower left. In this same area of the output map (Fig. 13, right) the vast majority of sources are classified as $n_{FG} = 20$ sources. We also see $n_{FG} = 20$ outside of the lobe dominated by FG in the training set map. The area dominated by $n_{RGB} = 20$ sources blends into the region in which $n_{FG} = 20$ sources appear due to the contamination between these classes (Sect. 4.2.2). AGN occupy several locations within the extragalactic regions of the t-SNE map for the training set, perhaps unsurprising given their wide range of physical properties, however most AGN are concentrated on an island above the tip of the AGB spur. In the t-SNE map for classified sources however the $n_{AGN} = 20$ sources are tightly concentrated on that same location with no scattered points; this is a likely consequence of our PRF classification methodology since here we plot only $n_{AGN} = 20$ sources and these are more likely to be those which match the bulk of the training set in parameter space.

Based on Fig. 13, it is clear that an unsupervised machine learning method like the t-SNE using near-IR and far-IR features can be used to identify some classes of objects like RSG, OAGB and CAGB. Even though YSOs do cluster in such diagrams, such clusters are distributed across the map. Therefore, using the t-SNE maps to identify YSOs would be more difficult without any additional information and thus inherently less reliable. Due to the nature of t-SNE maps the positions of classes shown here apply within the constraints of this data set and the input parameters used (Sect. 3.1).

## 5 DISCUSSION

### 5.1 The YSO population of NGC 6822

The PRF identifies 368 sources with $n_{YSO} > 0$; of these 269 have $n_{YSO}$ as their largest $n_{class}$ value. We break these sources down into two categories based on the number of PRF runs in which they are identified as a YSO. There are 182 sources identified as YSOs in all the PRF runs ($n_{YSO} = 20$); these are classified as highest probability YSOs. A further 87 sources have $20 > n_{YSO} > 10$, and are classed as (probable) YSO candidates.

The catalogue of 324 YSOs and YSO candidates (including the 55 sources from the training set extension) is provided in Table 3; 173 sources were identified for the first time using our PRF methodology, 111 YSOs and 62 YSO candidates, as described below. In the follow-ing subsections we compare this YSO catalogue to those in previous works and comment on the physical properties of the sources.

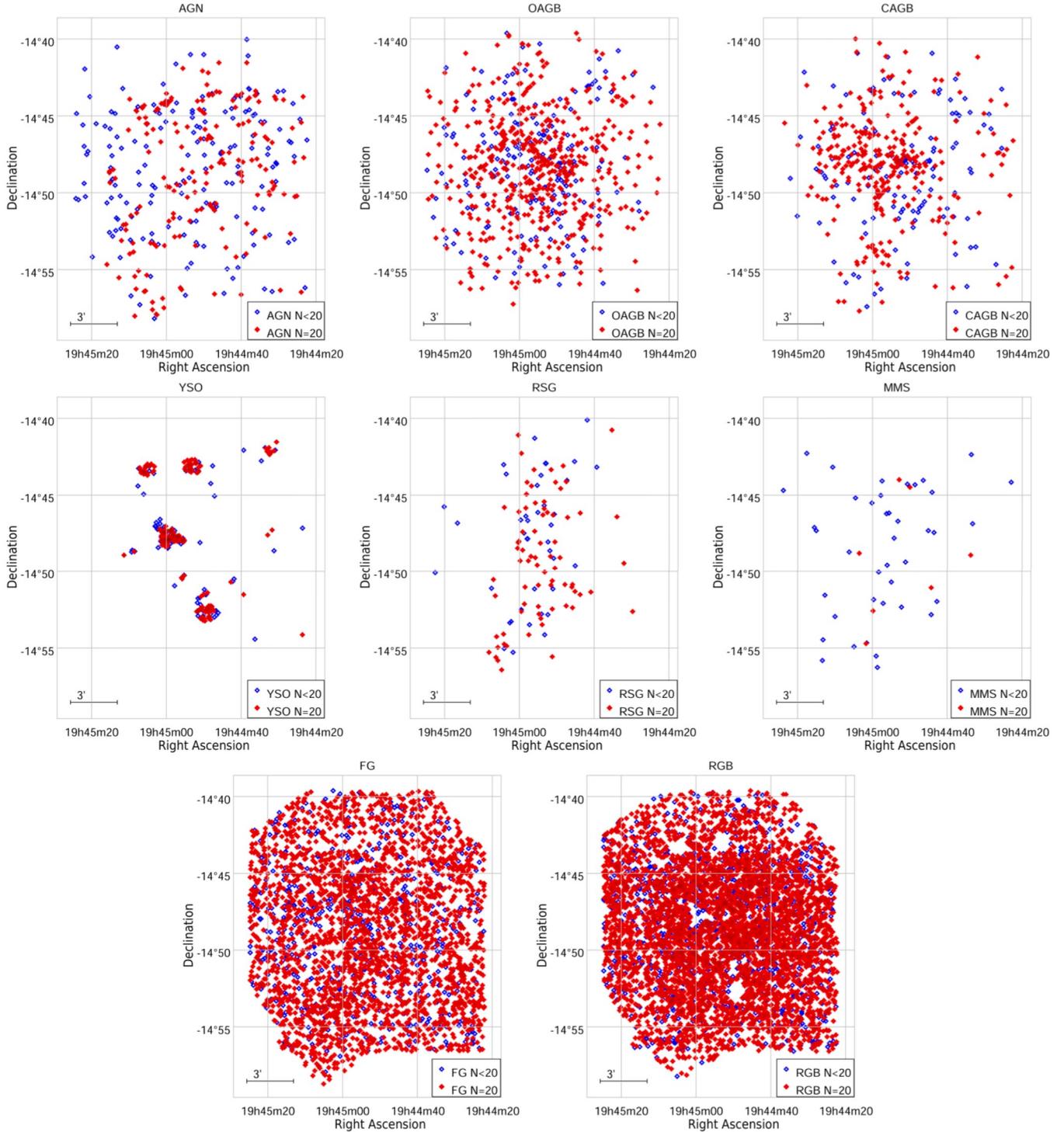

**Figure 11.** Spatial distributions on the sky of each target class from our enhanced classification. Sources with $n_{\mathrm{class}} = 20$ and $10 < n_{\mathrm{class}} < 20$ are shown in filled red and open blue diamonds respectively.

### 5.1.1 The classifications of known YSOs

The YSOs candidates from the catalogues of Jones et al. (2019) and Hirschauer et al. (2020) with near-IR counterparts not included in the extended training set were part of the catalogue of sources to be classified. For these 222 sources, we examined their output classifications in greater detail; 126 sources have $n_{class} = 20$ and the remaining 96 sources have $n_{class} < 20$.

For the $n_{class} = 20$ sources over a third are classified by the PRF as YSOs (Fig.14, upper panel). This is followed by RGB, CAGB and FG classifications that together account for another third of the classifications. The lowest reliability YSOs from Jones et al. (2019) dominate most classes in Fig. 14 (shown in grey), unsurprising given that such sources vastly outnumber higher confidence YSOs. The highest reliability sources from Jones et al. (2019) with $n_{class} = 20$ are classified mainly into the YSO class, with a smaller number in the RGB, FG and AGN classes. Of the 126 $n_{class} = 20$ sources 7, 7, and 91 come from the high, medium and low reliability tables of Jones et al. (2019) respectively, the remainder are sources from Hirschauer et al. (2020). Overall ∼ 38 per cent of the sources identified as YSOs in either Jones et al. (2019) and Hirschauer et al. (2020) with near-IR counterparts are classified by the PRF into another class with high confidence, i.e. $n_{class} = 20$.

Table 3. Catalogue of YSOs and YSO candidates in NGC 6822 classified using the PRF analysis. For sources previously identified as YSOs, the reference is provided in the last column, either Jones et al. (2019, J19) or Hirschauer et al. (2020, H20). Sources included in the training set extension are marked with *. A sample of the table is provided here, the full catalogue is available as supplementary material.

| RA (J2000) h:m:s | Dec (J2000) deg:m:s | $J$ mag | $J_{err}$ mag | $H$ mag | $H_{err}$ mag | $K$ mag | $K_{err}$ mag | YSO status | Previous Identification |
|---|---|---|---|---|---|---|---|---|---|
| 19:45:04.36 | −14:43:04.9 | 18.41 | 0.070 | 17.84 | 0.052 | 17.55 | 0.059 | YSO | |
| 19:44:49.22 | −14:52:26.7 | 20.02 | 0.452 | 18.78 | 0.271 | 19.62 | 0.660 | YSO | H20 |
| 19:44:54.21 | −14:43:18.2 | 18.70 | 0.240 | 18.14 | 0.226 | 17.64 | 0.180 | YSO* | J19/H20 |
| 19:45:00.39 | −14:47:40.1 | 17.65 | 0.075 | 16.74 | 0.051 | 16.54 | 0.050 | YSO candidate | J19 |

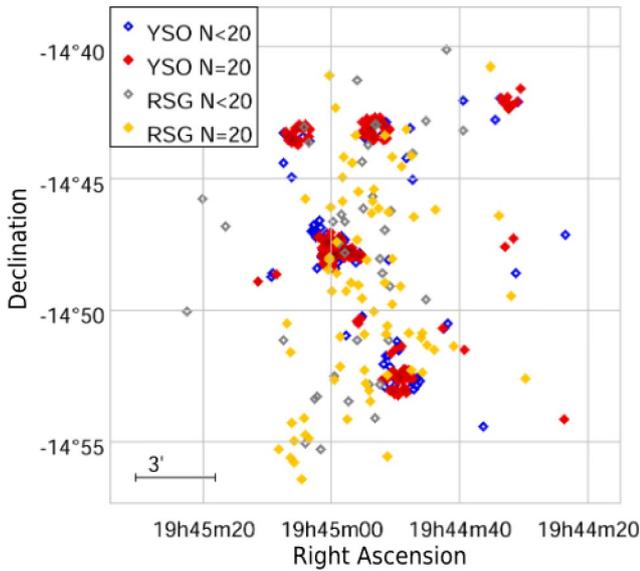

Figure 12. The spatial distribution on the sky of YSO and RSG sources. YSOs and RSGs with $n_{class} = 20$ are shown in red and gold respectively; sources with $10 < n_{class} < 20$ are shown in blue and grey, respectively.

The remaining 96 sources, with $10 < n_{class} < 20$, are classified into more than one class across the 20 PRF runs but have a majority consensus classification. In Fig. 14 (lower panel) we show the target class with the greatest number of classifications for each source. As above, sources from the low confidence table in Jones et al. (2019) are the most numerous. Most sources are classified as YSOs, followed by CAGBs and FGs. Sources from the highest confidence table of Jones et al. (2019) are categorised into YSO, CAGB, and AGN classes. The YSO candidates from Hirschauer et al. (2020) are classified as RGBs and FGs less often than those from Jones et al. (2019), however the opposite is true for the CAGB class.

From a total of 807 YSO candidates from Jones et al. (2019) and Hirschauer et al. (2020) we have sufficient feature information to classify 277 sources. We confirm the YSO nature for 125 of these (77 as YSOs and 48 as candidates) with 55 included in the training set extension and the remainder classified by the subsequent PRF runs: in detail 15/23 high-, 6/18 mid- and 76/195 low-reliability sources from Jones et al. (2019), and 28/41 sources from Hirschauer et al. (2020). Overall the confirmation rate of previously known YSO candidates is ∼ 44 per cent, however it is ∼ 65 per cent for the higher-reliability sample. Furthermore 82 out of the 277 literature YSO candidates are classified with high degree of confidence ($n_{class} = 20$) in the following PRF classes: seventeen FGs, twenty eight RGBs, eleven AGNs, seventeen CAGBs and nine OAGBs. There are a further two literature sources with no majority classification (no class has $n_{class} > 10$).

Of the 215 unique YSOs identified with the PRF (including training set extension sources), 76 and 31 sources were previously classified as YSOs respectively by Jones et al. (2019) and Hirschauer et al. (2020); of the 87 YSO candidates 42 and 15 were also identified in those papers. This accounts for ∼ 40 per cent of the YSOs and candidates we classify. Therefore we *classify for the first time* 199 sources, 136 of which are YSOs and 63 are candidates.

*5.1.2 YSO properties*

Jones et al. (2019) provide YSO masses and evolutionary stages derived using the SED models of Robitaille et al. (2006) and Robitaille (2017) for their high-confidence sample. All YSOs in common with our sample are best fitted by Stage I models (i.e. still relatively embedded). By comparing the $K$-band magnitude range of the PRF-identified YSOs with that of the YSO candidates from Jones et al. (2019) with mass determinations, we estimate a mass range for the newly identified YSOs between $15 − 50 \, M_\odot$. These massive YSOs are more likely the dominant source in an unresolved cluster (Oliveira et al. 2013; Ward et al. 2016, 2017). Indeed Jones et al. (2019) note the effect of multiplicity on a comparable YSO model fitting analysis from Chen et al. (2010) and hence present their mass estimates as overestimated for the dominant source but underestimated for the total unresolved cluster.

The CO (2−1) map of Gratier et al. (2010) covers the Northern section of the galaxy's bar (Fig. 1), with significant gaps in the coverage between the major SFRs. To explore the potential use of CO emission as a feature in the PRF identification of YSOs we perform large aperture photometry in the same way as described in Sect. 2.2 for the far-IR data.

CO brightnesses were measured for 1061 sources, 71 of which are YSOs and 30 YSO candidates (Fig. 15). YSOs exhibit on average higher CO brightness, with slightly lower average values seen for candidate YSOs. CO brightnesses for non-YSO sources are on average even lower: the median CO brightness values for YSOs and candidates are higher than that for non-YSOs by factors ∼ 2 and ∼ 1.5 respectively. We conclude that unresolved CO brightness can be a powerful discriminant between YSOs and other stellar populations over large areas.

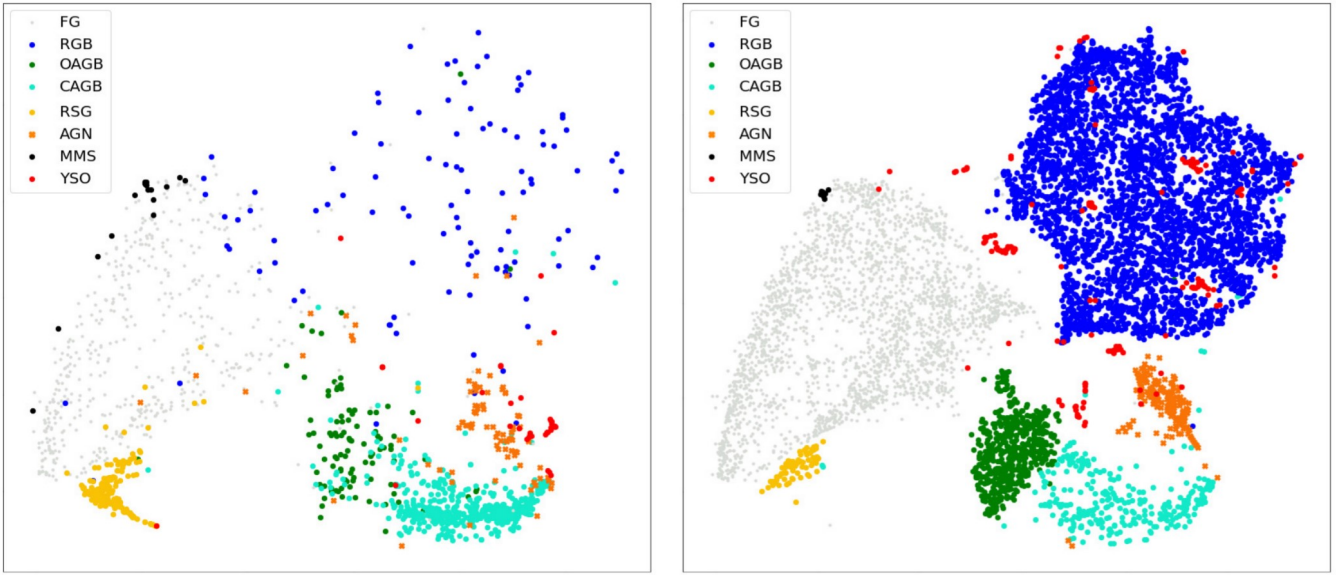

**Figure 13.** t-SNE maps for the training set data (left) and PRF classification outputs with $n_{class}$ = 20 (right), colour-coded as Figs. 3 and 10. Note that the axes for a t-SNE plot are unitless.

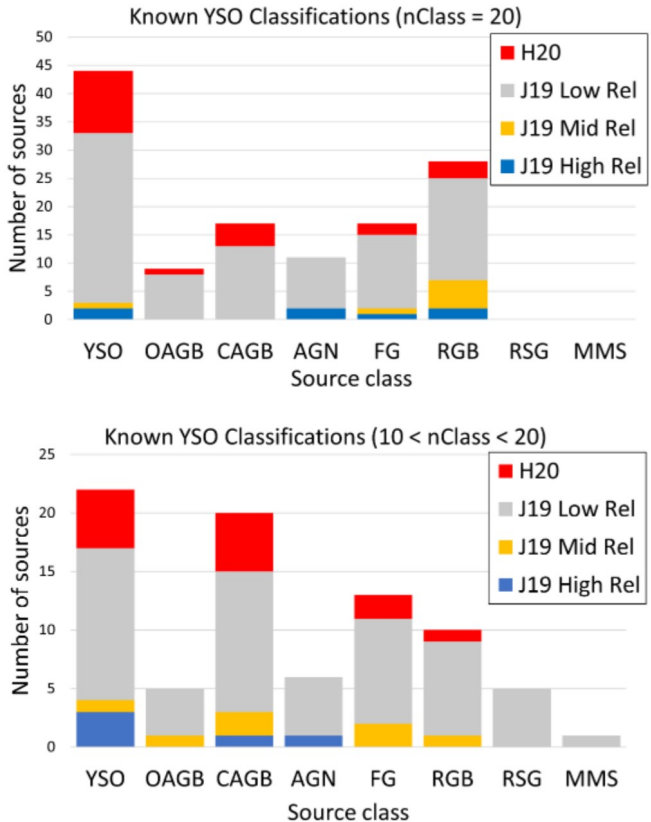

**Figure 14.** PRF classifications for previously known YSO candidates from Jones et al. (2019) and Hirschauer et al. (2020) with $n_{class}$ = 20 (top) and 10 < $n_{class}$ < 20 (bottom, showing the majority consensus classification). The reliability levels from Jones et al. (2019, J19) and YSO candidates unique to Hirschauer et al. (2020, H20) are colour-coded.

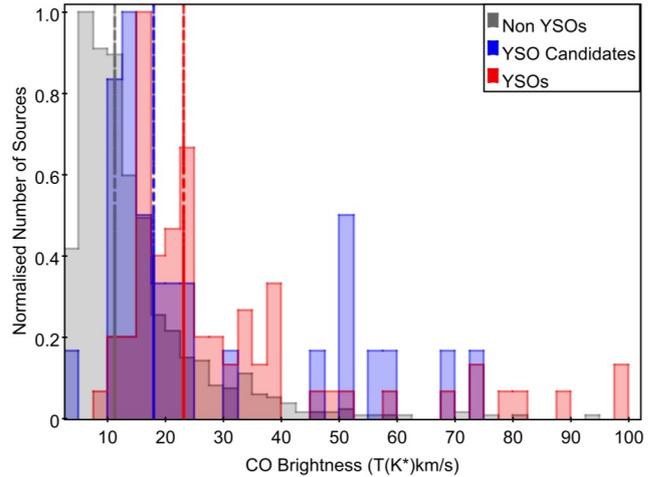

**Figure 15.** A normalised histogram of CO brightness for YSOs, candidates and non-YSO sources. The median value for each group is shown by the vertical dashed line of the same colour.

### 5.2 The star formation environment in NGC 6822

We classified YSOs and YSO candidates in all the major (known) SFRs shown in Fig. 1 as well as outside these regions in smaller numbers (Figs. 12 and 16). The number of YSOs and candidates classified in each of the SFRs is provided in Table 4. None of the YSOs or YSO candidates are coincident with the globular or young clusters in the region (see Introduction).

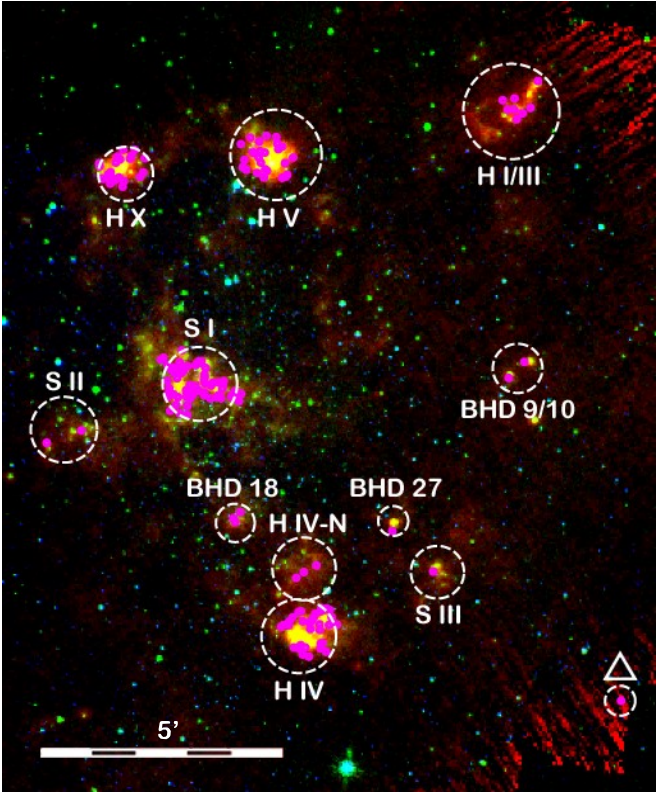

**Figure 16.** RGB image of NGC 6822 (respectively *Herschel* PACS 160 $\mu$m, *Spitzer* IRAC 8 $\mu$m and WFCAM $J$-band) with $n_{YSO}$ = 20 sources identified (magenta squares). The seven Hubble and Spitzer regions are shown with the radii given by Jones et al. (2019). The regions BHD 9/10, 18, 27 and Hubble IV-N are newly identified in this work as star formation sites. The region marked with an upright triangle shows the position of the single YSO discussed in the final paragraph of Sect. 5.2.

The ratio of H$\alpha$ emission from less embedded young stars to mid-IR emission from warm dust surrounding embedded sources is greater in older, more evolved regions as radiative feedback from massive young stars clears the interstellar medium. Using a comparison of CO and H$\alpha$ morphologies as well as the H$\alpha$/24 $\mu$m emission ratio, Schruba et al. (2017) suggest a more embedded stage of star formation for Hubble IV and V indicative of a younger age, while Hubble I/III and X, that present fewer signatures of embedded star formation (more dispersed morphologies and a higher H$\alpha$/24 $\mu$m), are more evolved.

Using a similar methodology as Schruba et al. (2017), Jones et al. (2019) suggest an evolutionary stage for Spitzer II and Spitzer III similar to that of Hubble IV and V, even though the former two look relatively inconspicuous in their fig. 10. They propose that Spitzer I is the youngest region due to faint H$\alpha$ and UV emission and strong mid-IR emission. Since many embedded YSOs are identified in this region they conclude that the rate of star formation is near its peak. Spitzer I is also the region with the highest number of YSOs in our catalogue. In Fig. 16 the green glow of 8 $\mu$m emission tracing warm dust is far greater compared to, for example, Hubble I/III which is thought to be the more evolved SFR in NGC 6822.

Since we find a low number of YSOs and candidates in Spitzer II (see Table 4), we discuss this region in a little more detail. In Spitzer II there are 30 literature YSO candidates of which all but four originate from the low reliability sample of Jones et al. (2019); the remainder are one high and one medium reliability sources from Jones et al. (2019) and two YSO candidates from Hirschauer et al. (2020). We find near-IR counterparts for eighteen literature YSO candidates of which fifteen are classified with $n_{class}$ = 20 by the PRF analysis, but only two are classified as YSOs (counterparts to one high- and one low-reliability YSO candidates, Jones et al. 2019). In addition we identity a further three YSO candidates (10 < $n_{YSO}$ < 20), two of which are in the low-reliability sample (Jones et al. 2019). The remaining $n_{class}$ = 20 sources in Spitzer II are classified as five RGBs, four OAGBs, two CAGBs and two FGs. Therefore, given that most literature YSO candidates were considered to be of low-reliability, it is not surprising that we only find two YSOs and three candidates in Spitzer II.

Using *Herschel* far-IR and *Spitzer* mid-IR data Galametz et al. (2010) found that region integrated SEDs between 10 − 100 $\mu$m show signatures of evolution for some of the H II regions. They propose that the 250 $\mu$m/500 $\mu$m emission ratio in particular correlates with the 24 $\mu$m emission and thus traces star formation activity. According to that ratio (see their figure 3b), Hubble V would be the most active region followed by Hubble IV, Spitzer I and Hubble X. These are the regions with the largest number of YSOs both in our analysis and that of Jones et al. (2019). Given that Hubble X and V have strong H$\alpha$ emission, they would have evolved past the peak of star formation activity. Spitzer I and Hubble IV on the other hand would be at their peak; for Spitzer I this is supported by the largest number of YSOs, however this is less clear for Hubble IV.

Looking at the positions within the SFRs of the PRF-classified YSOs, in Hubble IV and V the literature YSOs are more centrally concentrated. This is likely due to limitations in recovering point sources within the centre of these bright SFR in the near-IR images (see Sect. 2.1.2). In the other regions there is no significant difference in the locations of the YSOs within SRFs. In Hubble I/III the PRF classifies YSOs primarily in Hubble I and at the interface of the regions, in agreement with Jones et al. (2019) who suggest that Hubble I is more actively forming stars.

One of the two unnamed clusterings of YSOs noted (but not discussed in detail) in Jones et al. (2019) and recovered in our classification is coincident with a pair of H II regions; the first was designated K$\alpha$ (Kinman et al. 1979) and the second HK1 (Hodge et al. 1988). These regions are listed in the catalogue of Brunthaler et al. (2006) as BHD 9 and 10; they are located around 19:44:32, −14:47:26, almost directly South of Hubble I/III and to the West of the bar, away from the bulk of star formation activity (Fig. 16). The other unnamed region is found to the North-East of Spitzer III (at 19:44:42, −14:50:39) within 10 arcsec of the H II region BDH 27 (Brunthaler et al. 2006), in which we detect a single YSO and two candidates. In a region in the bar approximately equidistant from Spitzer I and Hubble IV (at 19:44:55, −14:50:24) we find three YSOs and one candidate. This area is bright at 8 $\mu$m and in H I emission. The candidate YSO is coincident with a very bright H$\alpha$ point source or small bubble (BHD 18, Brunthaler et al. 2006). These three regions are clearly seen in the 250 $\mu$m/500 $\mu$m ratio map of Galametz et al. (2010), suggesting star formation activity is present. We label these newly identified regions of star formation using the Brunthaler et al. (2006) denominations (BHD 9/10, 18, and 27) in Fig. 16 and list the number of YSOs in each region in Table 4.

We classify three YSOs and six candidates directly North of Hubble IV but outside the SFR radius defined by Jones et al. (2019). The YSOs trace a line at the centre of this region which we name as Hubble IV-N (located at 19:44:50.00 −14:51:31.0). Hubble IV-N is bright in both 8 and 160 $\mu$m emission (see Fig. 16) but compara-tively faint at 24 and 70 $\mu$m. This region has no visible large-scale H$\alpha$ emission and it is also relatively bright in the 250 $\mu$m/500 $\mu$m ratio map of Galametz et al. (2010). The location of these newly identified sources along with that of those found in BHD 18 are very suggestive of additional star-formation activity in the bar of NGC 6822 between the major regions Hubble IV and Spitzer I. Given that our analysis deals only with the most massive YSOs (15 − 50 M$_\odot$, see Sect. 5.1.2) there is potential for more, lower-mass, YSOs to be found in this bar region.

**Table 4.** The number of YSOs ($n_{YSO} = 20$), candidate YSOs ($10 < n_{YSO} < 20$), and training set extension YSO sources (see Sect. 4.1.2) classified in each of the previously known SFRs in NGC 6822, as well as in newly identified YSO groupings (see discussion in the text). The extent of the major SFRs was taken from table 9 in Jones et al. (2019), and is shown in Fig. 16.

| SFR | YSO number | YSO candidate number | Training Set Extension YSOs |
|---|---|---|---|
| Hubble I/III | 9 | 4 | 1 |
| Hubble IV | 35 | 14 | 6 |
| Hubble V | 36 | 11 | 13 |
| Hubble X | 29 | 5 | 5 |
| Spitzer I | 90 | 49 | 26 |
| Spitzer II | 2 | 3 | 0 |
| Spitzer III | 2 | 0 | 1 |
| BHD 9/10 | 4 | 1 | 2 |
| BHD 18 | 3 | 0 | 1 |
| BHD 27 | 1 | 2 | 0 |
| Hubble IV-N | 3 | 6 | 0 |

There is a YSO to the South-West of our field with $n_{YSO} = 20$. At this location, there is no UV emission in the images of Hunter et al. (2010), but the H$\alpha$ image from Massey et al. (2007) shows a point source. Mid-infrared emission is also unremarkable with a point-source source visible at 3.6 $\mu$m but not at 8 $\mu$m. Far-IR emission is not prominent but this location is close to the edge of these images. This source is identified with △ in Fig. 16 (located at 19:44:23.64, −14:54:07.9). This source could represent an isolated YSO, perhaps at the lower mass limit of our current detection range.

In addition, there are a further thirteen isolated YSO candidates located outside the SFRs in Fig. 16. These candidates are less certain ($n_{YSO} < 15$), and thus we do not discuss them further.

## 6 CONCLUSIONS

With a combination of near-IR and far-IR features we have used machine learning algorithms based on a probabilistic random forest classifier (PRF) and t-distributed stochastic neighbour embedding (t-SNE) to classify stellar populations in the main bar of NGC 6822, covering all prominent SFRs.

The PRF was trained using three near-IR colours ($J − H$, $H − K$ and $J − K$), $K$-band magnitude and two far-IR brightnesses (at 70 and 160 $\mu$m) and classifies sources into eight target classes (YSO, OAGB, CAGB, RGB, RSG, MMS, FG and AGN) with an estimated accuracy of 91 per cent across all classes rising to 96 per cent for YSOs (based on the PRF confusion matrices of the test sample).

We used the same near- and far-IR features to construct (unsupervised) t-SNE maps to identify stellar populations. Such maps are very effective in picking AGB stars (with a clear differentiation between OAGBs and CAGBs), AGN and RSG stars. Without additional information, the t-SNE maps seem however less powerful in identifying other classes of sources, including YSOs.

The spatial distributions of most stellar populations are essentially as expected. RSG stars, that trace the recent star formation history, occupy the bar of NGC 6822, linking the more conspicuous SFRs. An extension of the bar to the South-East, into a region which has indicators of youth (e.g. De Blok & Walter 2000) is seen in the RSG distribution, however no YSOs or candidates are classified there.

We classify a total of 324 YSOs and candidates. We confirm the nature of 125 out of 277 literature YSO candidates with enough feature information. Additionally 136 YSOs and 63 YSO candidates are classified for the first time in our analysis. We have not imposed a requirement that YSOs and candidates need to be located in main SRFs (as was done in previous works), and have detected YSOs in the bar of NGC 6822 between the major SFRs. YSOs classified by the PRF have mass estimates between ∼ 15 − 50 M$_\odot$, representing the most massive YSO population in NGC 6822. Another 82 out of 277 literature YSO candidates are definitively classified as non-YSOs by the PRF analysis.

We have identified YSOs in all known major star formation complexes in NGC 6822 (Hubble I/III, Hubble IV, Hubble V, Spitzer I, Spitzer II and Hubble X), but also in smaller star formation sites: the H<sub>II</sub> regions BHD 9/10 and 27 (noted but not analysed in Jones et al. 2019), as well as new regions of star formation BHD 18 and a region to the North but physically distinct from Hubble IV, that we name Hubble IV-N. The detection of massive YSOs in new regions, especially in BHD 18 and Hubble IV-N, is very suggestive of additional star formation occurring in the bar of NGC 6822 between the major previously known SFRs. The prospect of detecting further YSOs in the bar region in the mass regime below the sensitivity of our analysis remains to be explored.

Machine learning methods to classify large IR datasets will become increasingly important as the next generation of observatories such as the Extremely Large Telescope, *James Webb* and *Roman Space Telescopes* come online in the next decade. These new facilities will transform the range of galaxy distances in which resolved star formation studies are feasible, and increase the sensitivity to lower-mass and more-evolved YSOs in NGC 6822 and other Local Group galaxies. Machine learning techniques will prove invaluable in exploring such treasure trove of new data.

## ACKNOWLEDGEMENTS

The authors would like to thank the anonymous referee for their helpful comments and suggestions.

The authors thank W.J.G. de Blok, A. Schruba, P. Gratier and M. Irwin for their assistance with images used in this paper, I. Reis for help with PRF implementation, and O. Jones, A. Hirschauer and C. Pennock for access to their catalogues prior to full publication.

DAK acknowledges financial support from STFC via their PhD studentship programmes.

*Herschel* is a European Space Agency (ESA) space observatory with science instruments provided by European-led Principal Investigator consortia and with important participation from NASA.

This work has made use of data from the ESA mission *Gaia* (https://www.cosmos.esa.int/gaia), processed by the *Gaia* Data Processing and Analysis Consortium (DPAC, https://www.cosmos.esa.int/web/gaia/dpac/consortium). Funding for the DPAC has been provided by national institutions, in particular the institutions participating in the *Gaia* Multilateral Agreement.

# 7 DATA AVAILABILITY

The data underlying this article which are not included in the supplementary materials will be shared on reasonable request to the corresponding author.

# REFERENCES

Besla G., 2015, The Orbits of the Magellanic Clouds. Springer International Publishing, doi:10.1007/978-3-319-10614-4_26
Bianchi L., Scuderi S., Massey P., Romaniello M., 2001, AJ, 121, 2020
Bianchi L., Efremova B., Hodge P., Massey P., Olsen K. A. G., 2012, AJ, 143, 74
Bradley L., et al., 2020, astropy/photutils: 1.0.0, doi:10.5281/zenodo.4044744, https://doi.org/10.5281/zenodo.4044744
Breiman L., 2001, Machine learning, 45, 5
Brunthaler A., Henkel C., de Blok W. J. G., Reid M. J., Greenhill L. J., Falcke H., 2006, A&A, 457, 109
Cannon J. M., et al., 2006, ApJ, 652, 1170
Cannon J. M., et al., 2012, ApJ, 747, 122
Casali M., et al., 2007, A&A, 467, 777
Castelli F., Kurucz R. L., 2003, in Piskunov N., Weiss W. W., Gray D. F., eds, Astronomical Society of the Pacific Vol. 210, Modelling of Stellar Atmospheres. p. A20 (arXiv:astro-ph/0405087)
Chandar R., Bianchi L., Ford H. C., 2000, AJ, 120, 3088
Chen C.-H. R., et al., 2010, The Astrophysical Journal, 721, 1206
Cioni M. R. L., Habing H. J., Israel F. P., 2000, A&A, 358, L9
Clark C. J. R., Roman-Duval J. C., Gordon K. D., Bot C., Smith M. W. L., 2021, arXiv e-prints, p. arXiv:2107.14302
Cornu D., Montillaud J., 2020, arXiv e-prints, p. arXiv:2010.01601
De Blok W. J. G., Walter F., 2000, ApJ, 537, L95
De Blok W. J. G., Walter F., 2003, MNRAS, 341, L39
Dobrzycki A., Macri L. M., Stanek K. Z., Groot P. J., 2003, AJ, 125, 1330
Efremova B. V., et al., 2011, ApJ, 730, 88
Flesch E. W., 2021, MILLIQUAS - Million Quasars Catalog, Version 7.2, p. arXiv:2105.12985
Gaia Collaboration Brown A. G. A., Vallenari A., Prusti T., de Bruijne J. H. J., Babusiaux C., Biermann M., 2020, arXiv e-prints, p. arXiv:2012.01533
Galametz M., et al., 2010, A&A, 518, L55
Geha M., et al., 2003, AJ, 125, 1
Girardi L., Groenewegen M. A. T., Hatziminaoglou E., da Costa L., 2005, A&A, 436, 895
Gordon K. D., et al., 2011, AJ, 142, 102
Gottesman S. T., Weliachew L., 1977, A&A, 61, 523
Gratier P., Braine J., Rodriguez-Fernandez N. J., Israel F. P., Schuster K. F., Brouillet N., Gardan E., 2010, A&A, 512, A68
Hernandez E. J., Srinivasan S., Marshall J., 2021, in American Astronomical Society Meeting Abstracts. p. 541.06
Hilditch R. W., Howarth I. D., Harries T. J., 2005, MNRAS, 357, 304
Hirschauer A. S., Gray L., Meixner M., Jones O. C., Srinivasan S., Boyer M. L., Sargent B. A., 2020, ApJ, 892, 91
Hodge P., Kennicutt R. C. J., Lee M. G., 1988, PASP, 100, 917
Hodgkin S. T., Irwin M. J., Hewett P. C., Warren S. J., 2009, MNRAS, 394, 675
Hony S., et al., 2011, A&A, 531, A137
Hubble E. P., 1925, ApJ, 62, 409
Hunter D. A., Elmegreen B. G., Ludka B. C., 2010, AJ, 139, 447
Huxor A. P., Ferguson A. M. N., Veljanoski J., Mackey A. D., Tanvir N. R., 2013, MNRAS, 429, 1039
Ivanov V. D., et al., 2016, A&A, 588, A93
Jones O. C., et al., 2017, MNRAS, 470, 3250
Jones O. C., Sharp M. J., Reiter M., Hirschauer A. S., Meixner M., Srinivasan S., 2019, MNRAS, 490, 832
Kacharov N., Rejkuba M., Cioni M. R. L., 2012, A&A, 537, A108
Kato D., et al., 2007, PASJ, 59, 615
Kennicutt R. C. J., et al., 2003, PASP, 115, 928
Kinman T. D., Green J. R., Mahaffey C. T., 1979, PASP, 91, 749
Kirby E. N., Cohen J. G., Guhathakurta P., Cheng L., Bullock J. S., Gallazzi A., 2013, ApJ, 779, 102
Kozłowski S., Kochanek C. S., Udalski A., 2011, ApJS, 194, 22
Leisy P., Dennefeld M., Alard C., Guibert J., 1997, A&AS, 121, 407
Leisy P., Corradi R. L. M., Magrini L., Greimel R., Mampaso A., Dennefeld M., 2005, A&A, 436, 437
Letarte B., Demers S., Battinelli P., Kunkel W. E., 2002, AJ, 123, 832
Madden S. C., et al., 2014, PASP, 126, 1079
Maitra C., Haberl F., Ivanov V. D., 2018, in 42nd COSPAR Scientific Assembly. pp E1.12–27–18
Massey P., 1998, ApJ, 501, 153
Massey P., Olsen K. A. G., Hodge P. W., Jacoby G. H., McNeill R. T., Smith R. C., Strong S. B., 2007, AJ, 133, 2393
Mateo M. L., 1998, ARA&A, 36, 435
McConnachie A. W., Higgs C. R., Thomas G. F., Venn K. A., Côté P., Battaglia G., Lewis G. F., 2021, MNRAS, 501, 2363
Meixner M., et al., 2006, AJ, 132, 2268
Meixner M., et al., 2013, AJ, 146, 62
Mosteller F., Tukey J. W., 1968, in Lindzey G., Aronson E., eds, Handbook of Social Psychology, Vol. 2. Addison-Wesley
Neugent K. F., Levesque E. M., Massey P., Morrell N. I., Drout M. R., 2020, ApJ, 900, 118
Oliveira J. M., et al., 2013, MNRAS, 428, 3001
Pedregosa F., et al., 2011, Journal of Machine Learning Research, 12, 2825
Pietrzyński G., et al., 2013, Nature, 495, 76
Pilbratt G. L., et al., 2010, A&A, 518, L1
Poglitsch A., et al., 2010, A&A, 518, L2
Price-Whelan A. M., et al., 2018, The Astronomical Journal, 156, 123
Reis I., Baron D., Shahaf S., 2019, AJ, 157, 16
Rémy-Ruyer A., et al., 2015, A&A, 582, A121
Richer M. G., McCall M. L., 2007, ApJ, 658, 328
Rieke G. H., Lebofsky M. J., 1985, ApJ, 288, 618
Rieke G. H., et al., 2004, ApJS, 154, 25
Robitaille T. P., 2017, A&A, 600, A11
Robitaille T. P., Whitney B. A., Indebetouw R., Wood K., Denzmore P., 2006, ApJS, 167, 256
Schruba A., et al., 2017, ApJ, 835, 278
Sewiło M., et al., 2013, ApJ, 778, 15
Sibbons L. F., Ryan S. G., Cioni M. R. L., Irwin M., Napiwotzki R., 2012, A&A, 540, A135
Sibbons L. F., Ryan S. G., Napiwotzki R., Thompson G. P., 2015, A&A, 574, A102
Skillman E. D., Terlevich R., Melnick J., 1989, MNRAS, 240, 563
Swan J., Cole A. A., Tolstoy E., Irwin M. J., 2016, MNRAS, 456, 4315
Tan J. C., Beltrán M. T., Caselli P., Fontani F., Fuente A., Krumholz M. R., McKee C. F., Stolte A., 2014, in Beuther H., Klessen R. S., Dullemond C. P., Henning T., eds, Protostars and Planets VI. p. 149 (arXiv:1402.0919), doi:10.2458/azu_uapress_9780816531240-ch007
Tolstoy E., Irwin M. J., Cole A. A., Pasquini L., Gilmozzi R., Gallagher J. S., 2001, MNRAS, 327, 918
Van Loon J. T., 2008, Mem. Soc. Astron. Italiana, 79, 412
Van Loon J. T., Sansom A. E., 2015, MNRAS, 453, 2341
Van Loon J. T., Oliveira J. M., Gordon K. D., Sloan G. C., Engelbracht C. W., 2010, AJ, 139, 1553
Van der Maaten L., Hinton G., 2008, Journal of Machine Learning Research, 9, 2579
Volders L. M. J. S., Högbom J. A., 1961, Bull. Astron. Inst. Netherlands, 15, 307
Ward J. L., Oliveira J. M., van Loon J. T., Sewiło M., 2016, MNRAS, 455, 2345
Ward J. L., Oliveira J. M., van Loon J. T., Sewiło M., 2017, MNRAS, 464, 1512
Weldrake D. T. F., de Blok W. J. G., Walter F., 2003, MNRAS, 340, 12
Whitney B. A., et al., 2008, AJ, 136, 18
Zombeck M. V., 2006, Handbook of Space Astronomy and Astrophysics, 3 edn. Cambridge University Press, doi:10.1017/CBO9780511536359

# APPENDIX A: ON-LINE MATERIALS

Full versions of Tables 1 and 3 are available as supplementary material.

**Table 1.** Table of training set sources.
**Table 3.** Table of YSOs and YSO candidates.

# APPENDIX B: GAIA PROPER MOTIONS FOR THE FG AND MMS TRAINING SETS

Using the *Gaia* EDR3 data the near-IR sources were matched to proper motion (PM) information with a 1 arcsec matching radius. The distribution of PMs were analysed in both right ascension (RA) and declination (Dec) components separately in order to disentangle sources in NGC 6822 from Galactic foreground objects (Sect. 3.2). To achieve this we placed conservative limits on the PM component values as shown in Fig. B1; these limits are intended to obtain *clean samples* of FG and MMS sources rather than *complete samples*. We also compared the PM distributions in sources in the direction of NGC 6822 sources with those of two neighbouring off-galaxy areas of the same size as our target field to the North and South. The two off-field regions extend from 19:44:21 to 19:45:26 in right ascension. In declination the Northern field runs from −14:20:00 to −14:39:30 and the Southern from −14:59:00 to −15:17:50. Using the PM histograms we set limits for inclusion into the MMS training set of between −2 and 2 mas/yr in both RA and Dec (Sect. 3.4.4). For FG inclusion the limits are outside the range of −3 and 3 mas/yr in RA and −5 and 3 mas/yr in Dec (Sect. 3.4.3).

# APPENDIX C: NEAR-INFRARED PHOTOMETRIC TRANSFORMATIONS

For the training set sources either within or behind the MC, we use IRSF near-IR photometry (see Sect. 2.1.1). The transformations applied to convert from the IRSF to the WFCAM photometric systems are:

$$K_{\mathrm{WFCAM}} = K_{\mathrm{IRSF}} - 0.014 \tag{C1}$$

$$(J - H)_{\mathrm{WFCAM}} = 0.923 \times (J - H)_{\mathrm{IRSF}} + 0.036 \tag{C2}$$

$$(H - K)_{\mathrm{WFCAM}} = (H - K)_{\mathrm{IRSF}} + 0.055 \times (J - K)_{\mathrm{IRSF}} - 0.04 \tag{C3}$$

These were obtained by using the conversions from IRSF to 2MASS and WFCAM to 2MASS available respectively in Kato et al. (2007) and Hodgkin et al. (2009).

# APPENDIX D: FAR-INFRARED IMAGE CALIBRATIONS

The 160 $\mu$m images of the MCs, obtained with *Hercshel* PACS (see Sect. 2.2), have residual (non-astrophysical) bias levels that needed to be corrected for before the large aperture brightness measurements could be performed. This is not surprising given the complexity and challenges of processing these datasets from very early in the *Herschel* mission (see Meixner et al. 2013 for full details, and more recently Clark et al. 2021). These zero-level corrections were taken from pixel value histograms for each image which are shown in Fig. D1 for the SMC and in Fig. D2 for the LMC.

For consistently, we checked whether the 70 $\mu$m *Spitzer* MIPS images of the SMC and LMC (see Sect. 2.2) also required any adjustments and we accordingly applied very small correction of −0.14 and −0.05 MJy sr$^{-1}$ respectively in each pixel. The 160 $\mu$m images required offsets of +4.50 and +8.25 MJy sr$^{-1}$ in each pixel for the SMC and LMC respectively. The NGC 6822 images did not require any such corrections.

# APPENDIX E: FULL PRF CONFUSION MATRICES

The full set of confusion matrices for all extended PRF runs are shown here. A representative example from a single PRF run for both normalised and un-normalised matrices was shown in Fig. 8. The accuracy scores returned by SKLEARN for these runs vary between 87 and 92 per cent with an average of 91 per cent.

By comparing between runs with different random seeds in the un-normalised confusion matrices, the variations arising from the random selection of training and test samples can be seen, e.g. in their raw numerical values (Fig. E1). The strong diagonal seen in the normalised confusion matrices (Fig. E2) is weaker in the un-normalised matrices (Fig. E1) as a result of the different sizes of each target class. These un-normalised matrices do however show how many sources of each class are included in the testing data for each PRF run.

The normalised matrices allow for a better assessment of the success of the classifications, by evening out different class sizes. Fig. E2 shows these normalised confusion matrices. A good recovery rate can be seen in all classes with the exception of RGB stars which have a significant level of confusion with Galactic contaminants due to how the FG training set is constructed (see Sect. 4.2.2). The AGN class suffers from confusion with CAGB and FG classes in many runs. This is due in part to their similarities in near-IR colour, but also because of the limited number of available AGN training sources for a class with a large range of possible parameter space, as discussed in Sect. 4.1.1.

This paper has been typeset from a T<sub>E</sub>X/LAT<sub>E</sub>X file prepared by the author.

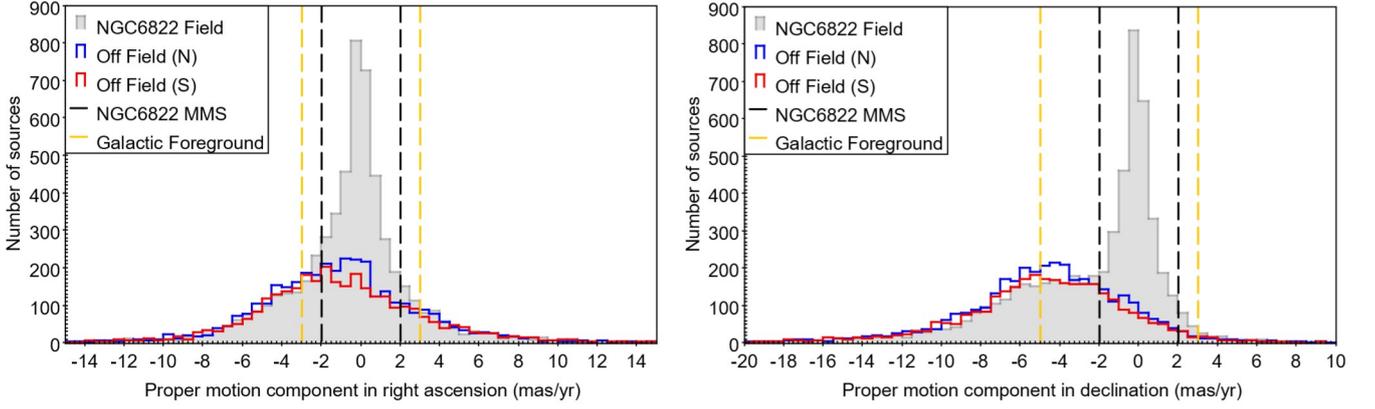

**Figure B1.** Histograms of proper motion components in RA and Dec with the limits for training set inclusion for MMS and FG classes shown. Off-galaxy comparison fields to the North (N) and South (S) are shown by the blue and red histograms respectively.

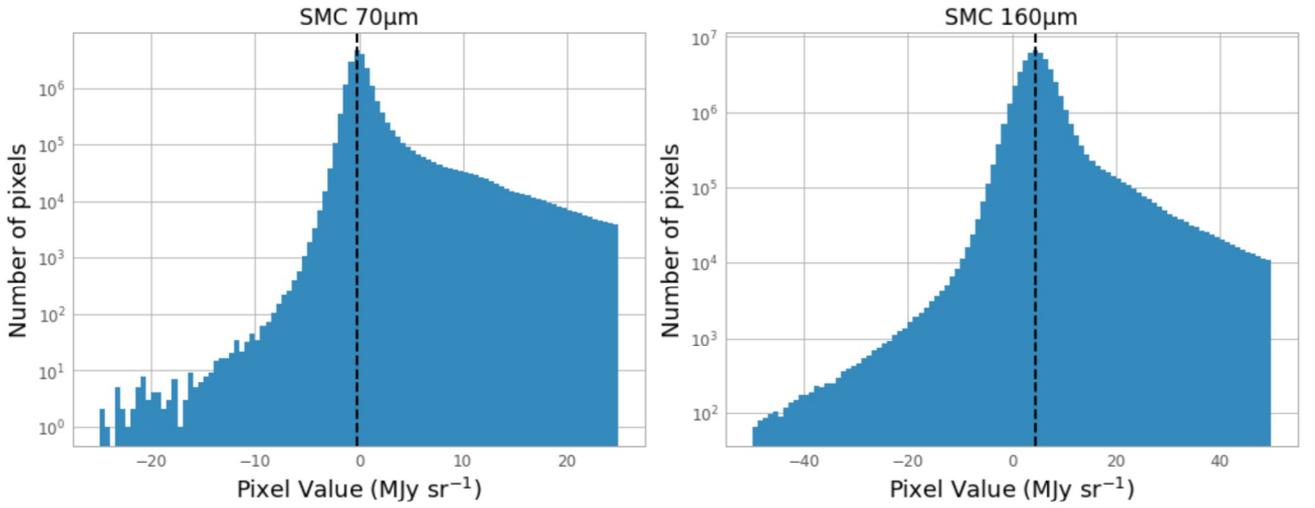

**Figure D1.** Histograms of the pixel values for the far-IR image of the SMC. Vertical dashed lines show the correction value applied.

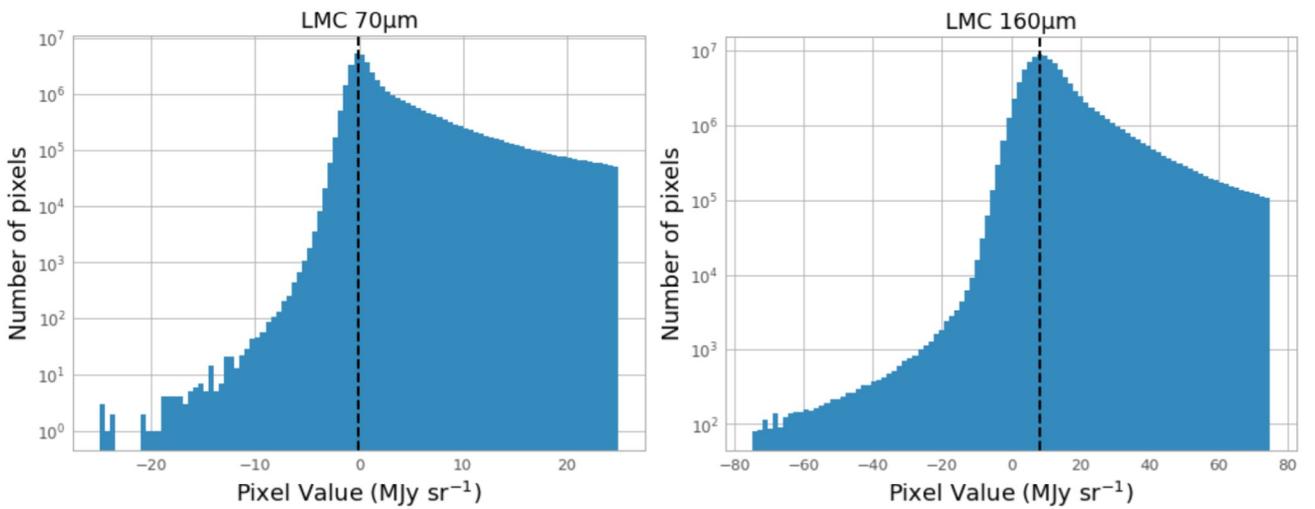

**Figure D2.** Histograms of the pixel values for the far-IR image of the LMC. Vertical dashed lines show the correction value applied.

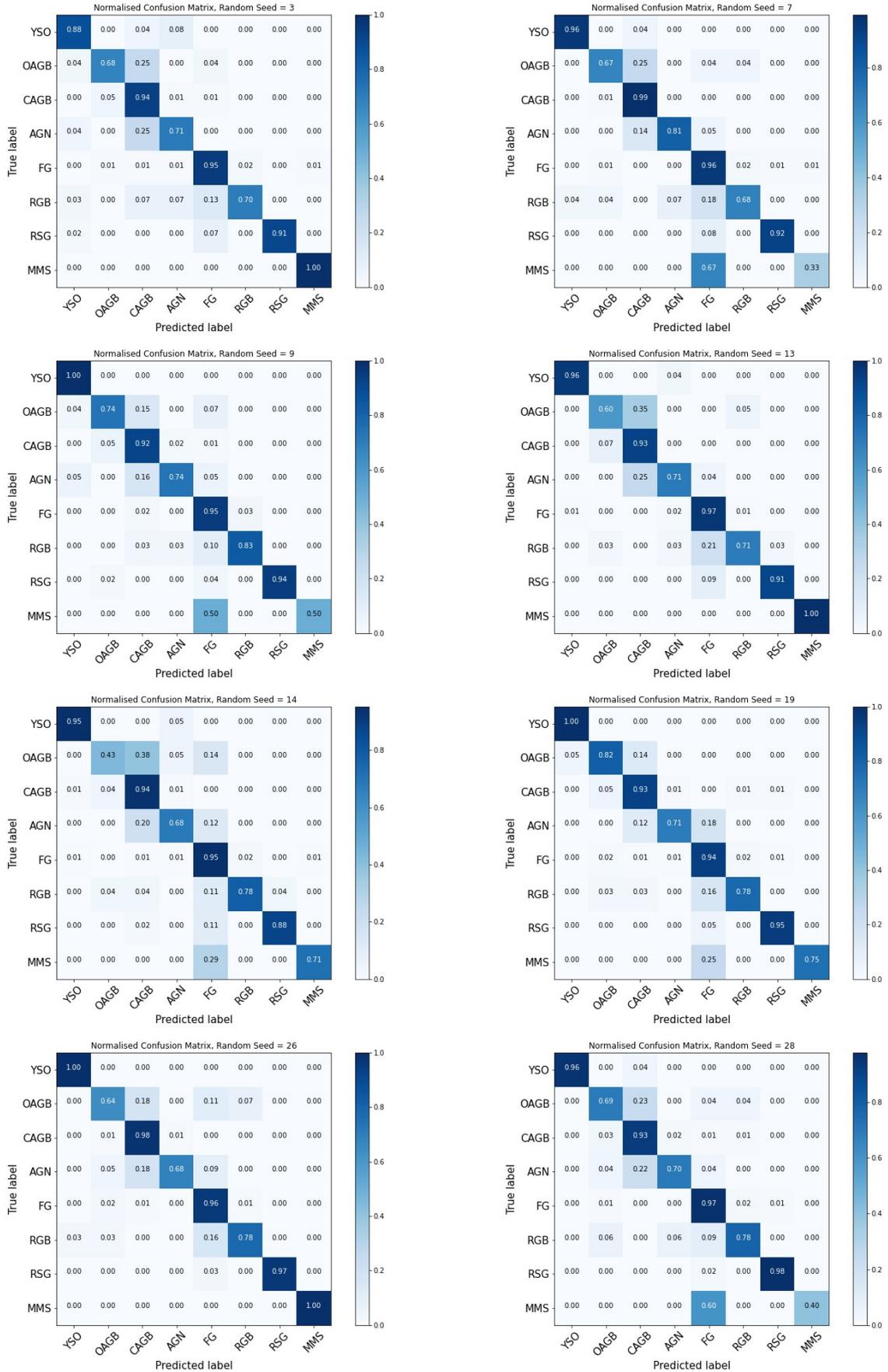

**Figure E1.** Confusion matrices for the 20 PRF runs using different random seeds to overcome any stochastic effects in train/test splitting.

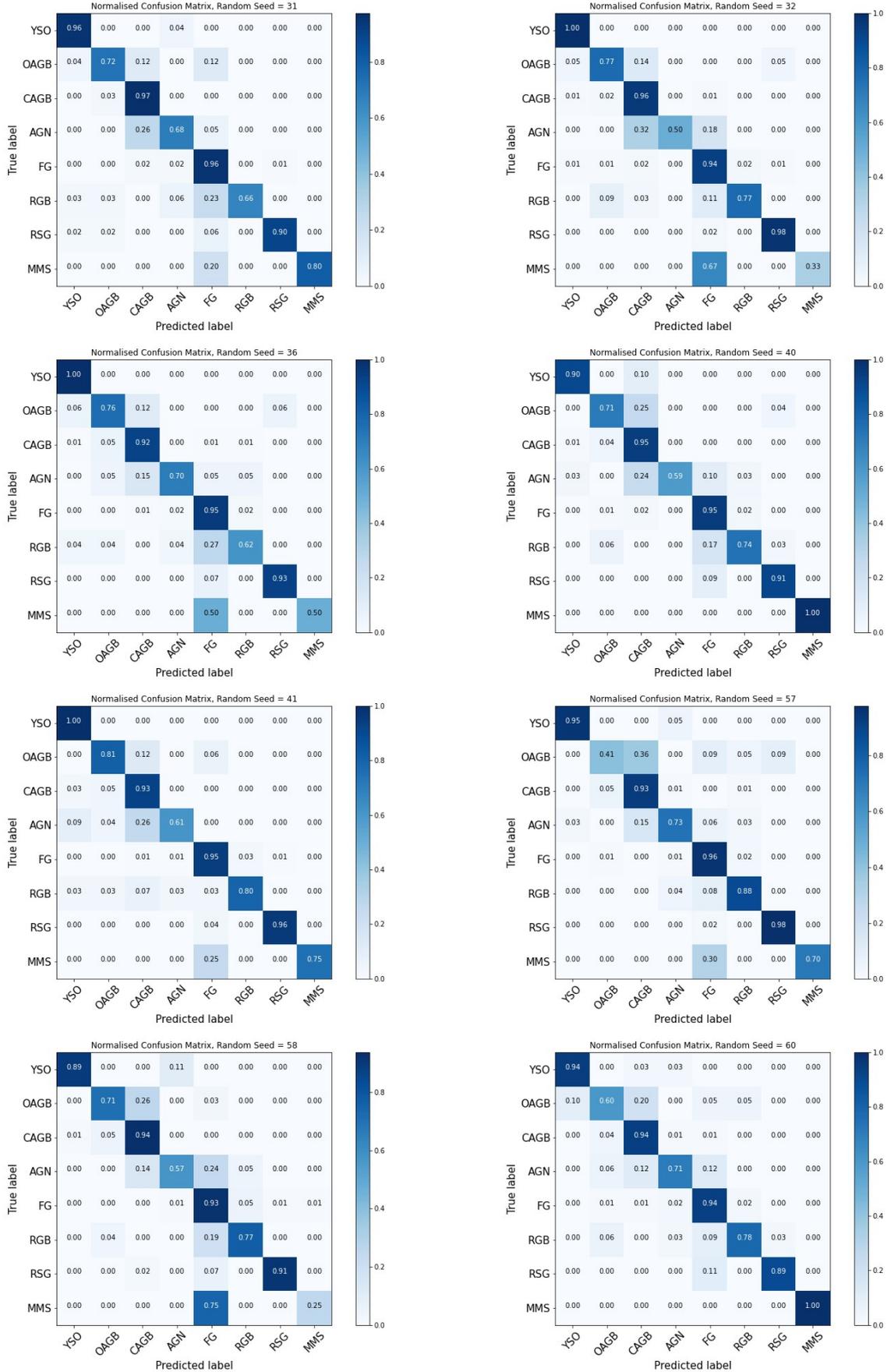

**Figure E1.** Cont.

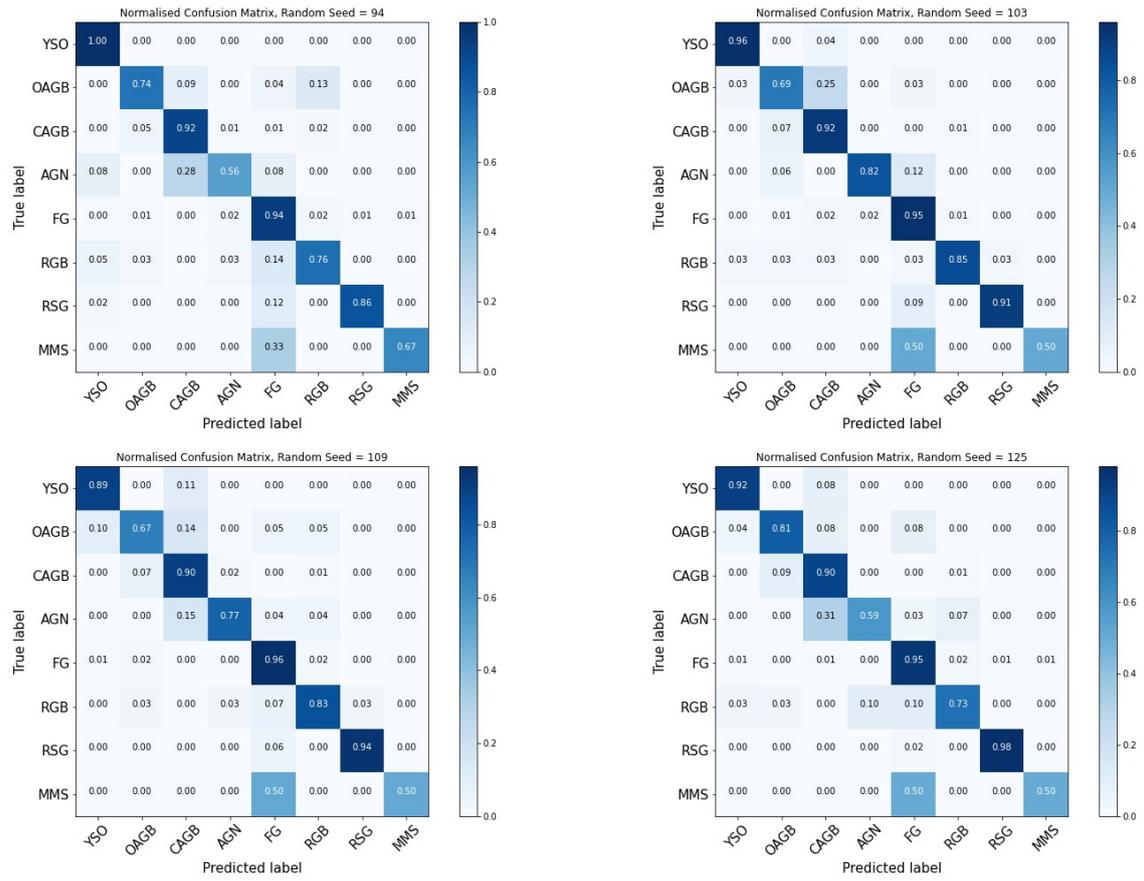

**Figure E1.** Cont.

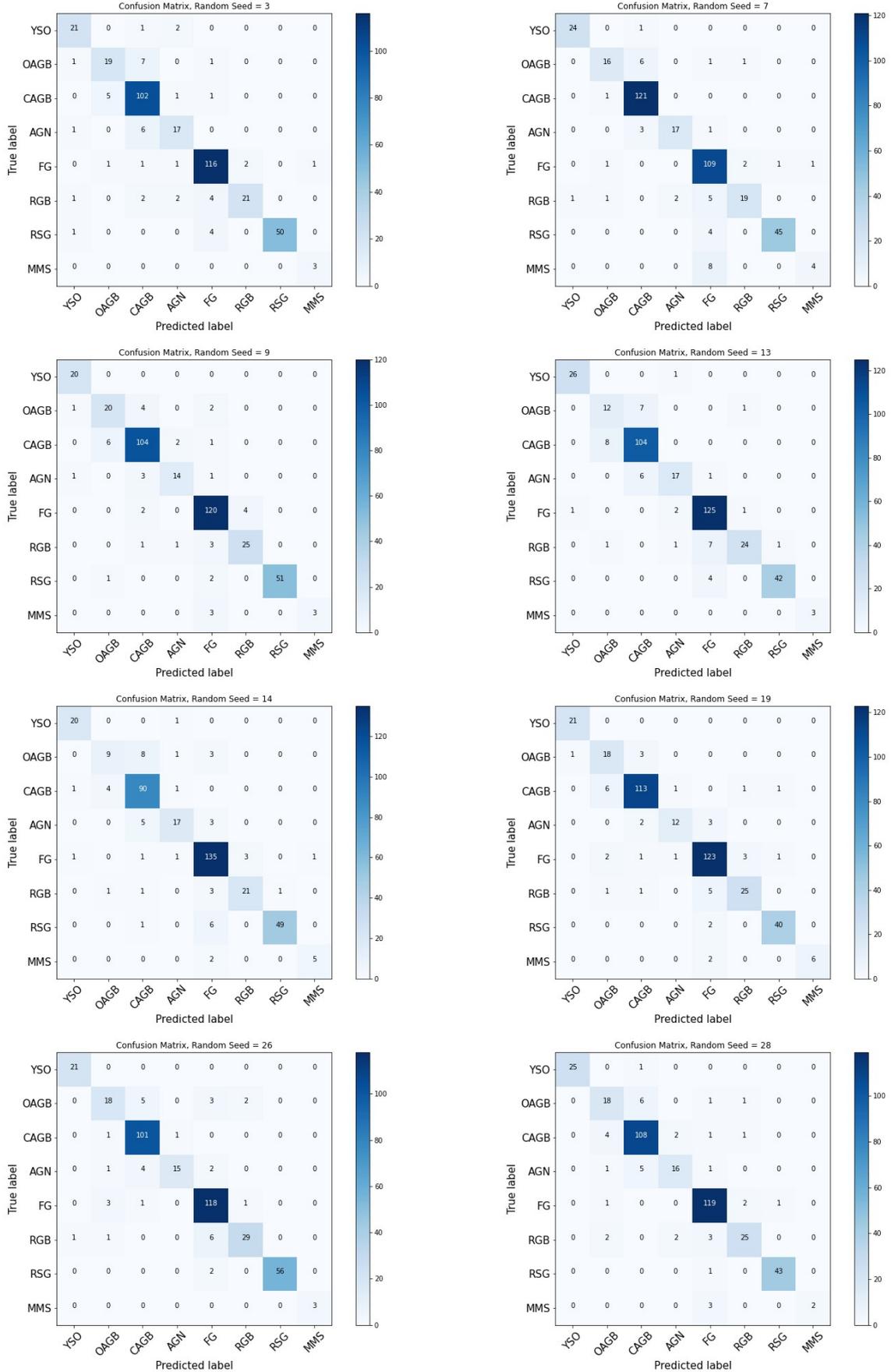

**Figure E2.** Normalised confusion matrices of the same runs shown in Fig.E1.

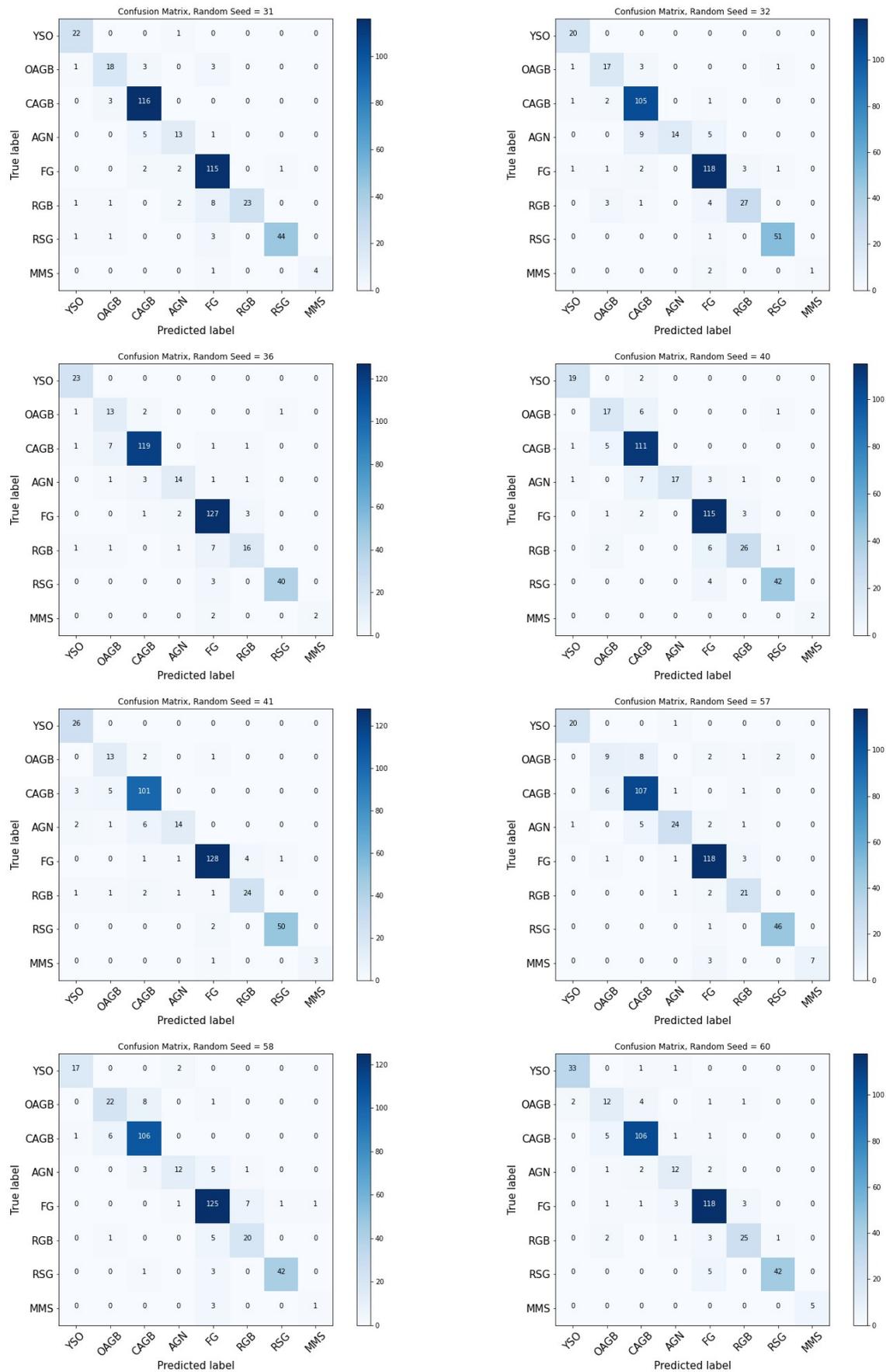

**Figure E2.** Cont.

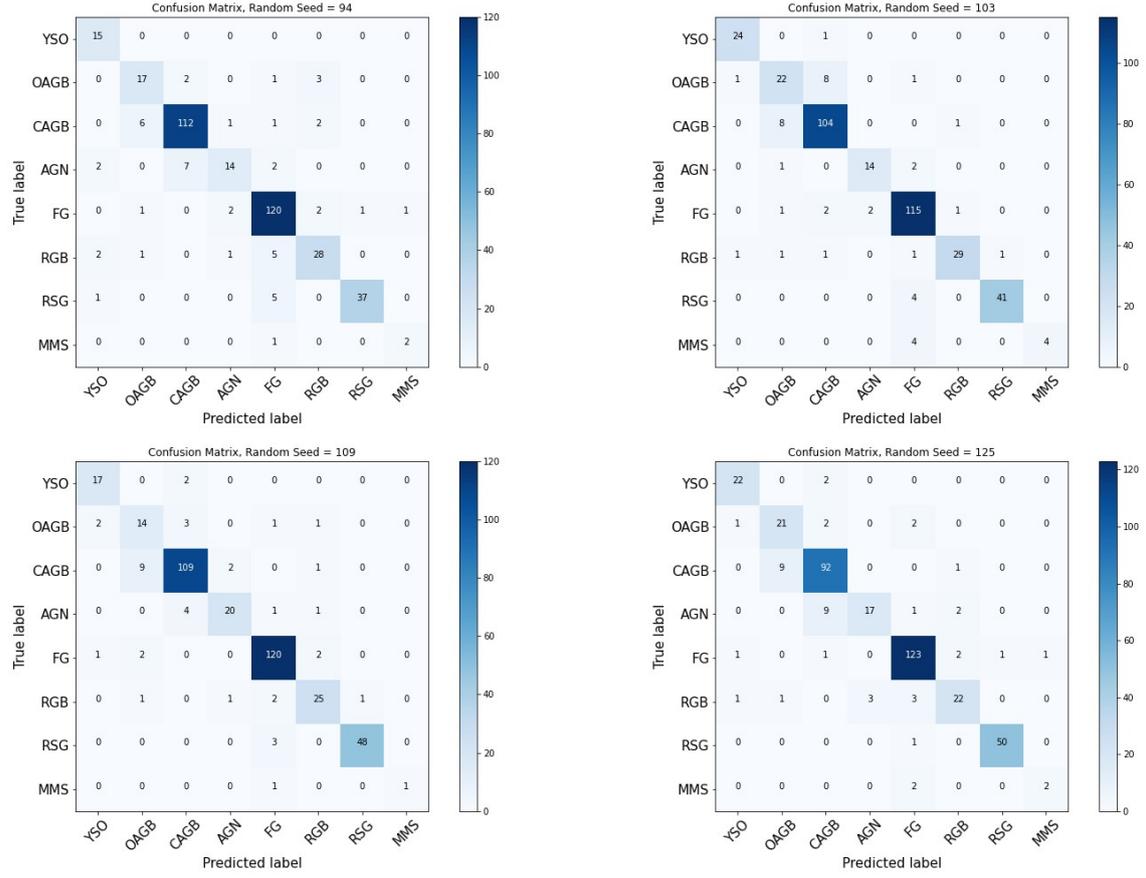

**Figure E2.** Cont.